%% file: arxiv.tex
\documentclass[a4paper,fleqn]{template/arxiv-cas-sc}

\usepackage[numbers]{natbib}

\usepackage{mathtools}
\usepackage{bm}
\usepackage{algorithm,algpseudocode}
\usepackage{amsthm}
\usepackage{subcaption}
\usepackage{tikz}
\usetikzlibrary{patterns}

\input{math_commands}

\input{sections/tikz_diagrams}

\begin{document}
\let\WriteBookmarks\relax
\renewcommand{\floatpagefraction}{0.5}
\renewcommand{\textfraction}{0.01}
\renewcommand{\topfraction}{0.95}
\renewcommand{\bottomfraction}{0.95}

\shorttitle{Schwarz Space-Time Refinement for MPM}
\shortauthors{Zhaofeng Luo, Minchen Li*, Yupeng Jiang*}

\title[mode=title]{An Overlapping Schwarz Space-Time Refinement Framework for Material Point Method}

\author[1]{Zhaofeng Luo}
\ead{zhaofen2@andrew.cmu.edu}

\author[1,2]{Minchen Li}
\cormark[1]
\ead{minchernl@gmail.com}

\author[3]{Yupeng Jiang}
\cormark[1]
\ead{yupeng.jiang@ibnm.uni-hannover.de}

\cortext[1]{Corresponding authors jointly supervised this work.}

\affiliation[1]{organization={Computer Science Department, Carnegie Mellon University},
            addressline={5000 Forbes Ave.},
            city={Pittsburgh},
            state={PA},
            postcode={15213},
            country={USA}}

\affiliation[2]{organization={Genesis AI},
            addressline={985 Industrial Rd.},
            city={San Carlos},
            state={CA},
            postcode={94070},
            country={USA}}

\affiliation[3]{organization={IBNM, Leibniz Universität Hannover},
            addressline={Welfengarten 1},
            city={Hannover},
            postcode={30167},
            country={Germany}}

\begin{abstract}
We propose an overlapping Schwarz space-time refinement framework for the material point method (OS-MPM) to improve computational efficiency in problems with strongly localized deformation, contact, and large geometric nonlinearity. The method decomposes the domain into overlapping coarse and fine subdomains with heterogeneous spatial and temporal resolutions, while retaining standard MPM discretizations within each subdomain. Coarse–fine coupling is achieved through an MPM-specific Schwarz iteration combining mass-weighted spatial transmission and temporal interpolation for sub-cycling. In contrast to refinement strategies based on modified basis functions, transition kernels, or strongly enforced interface constraints, the proposed approach preserves the modular structure of standard MPM and shifts the coupling complexity to nonmatching-grid interface operators within the Schwarz alternating procedure. Numerical examples, including a gravity-driven cantilever beam, Hertzian contact, and an elastic inclusion problem, show that the method reproduces analytical or fine-resolution reference solutions with good accuracy and convergence behavior. In the inclusion benchmark, the proposed framework achieves comparable or slightly lower error than single-domain fine simulations at the finest tested resolutions, while reducing computational cost by up to 9.15 times. A three-dimensional folding example further demonstrates the generality of the framework. These results indicate that the proposed method provides an accurate, modular, and efficient route for local space-time refinement in MPM.
\end{abstract}


\begin{highlights}
\item A novel overlapping Schwarz space-time refinement framework is developed for MPM, enabling coarse–fine coupling with heterogeneous spatial and temporal resolutions while preserving standard subdomain MPM discretizations.
\item The method introduces an MPM-specific interface treatment based on mass-weighted spatial transmission and temporal interpolation for sub-cycling, avoiding modified basis functions and explicit interface constraints.
\item Numerical benchmarks show that the framework preserves the accuracy and convergence behavior of monolithic fine-resolution simulations while achieving substantial computational savings, with reported speedups of up to 9.15 times.
\end{highlights}

\begin{keywords}
Material point method \sep
Overlapping Schwarz method \sep
Multi-resolution coupling \sep
Space-time refinement \sep
Implicit time integration
\end{keywords}

\maketitle


\input{sections/introduction}
\input{sections/formulation}
\input{sections/method}
\input{sections/results}

\input{sections/conclusion}

\appendix
\input{sections/appendix}

\bibliographystyle{cas-model2-names}
\bibliography{references}

\clearpage

\end{document}

%% file: math_commands.tex
\usepackage{amsthm} 


%% file: sections/tikz_diagrams.tex
\newcommand{\figInclusionSetup}{%
\begin{figure}[htbp]
    \centering
    \begin{tikzpicture}[scale=3.5]
        \begin{scope}[xshift=0cm]
            \node[align=center] at (0.5, 1.1) {\small (a) Coarse Domain $\Omega_B$};
            \draw[step=0.05, gray!30, thin] (0,0) grid (1,1);
            \draw[thick, darkgray] (0,0) rectangle (1,1);
            \filldraw[fill=blue!15, draw=blue!80!black, thick, even odd rule]
                (0.1,0.1) rectangle (0.9,0.9)
                (0.5,0.5) circle (0.1);
            \node[blue!80!black] at (0.5, 0.8) {\footnotesize Matrix};
            \draw[<->, >=stealth, densely dashed] (0.5,0.5) -- (0.6,0.5) node[midway, below] {\tiny $r=0.1$};
            \fill[black] (0.5,0.5) circle (0.3pt);
        \end{scope}

        \begin{scope}[xshift=1.4cm]
            \node[align=center] at (0.5, 1.1) {\small (b) Fine Domain $\Omega_S$};
            \draw[dashed, gray] (0,0) rectangle (1,1);
            \begin{scope}
                \draw[step=0.02, gray!40, thin] (0.3,0.3) grid (0.7,0.7);
                \draw[thick, green!50!black] (0.3,0.3) rectangle (0.7,0.7);

                \filldraw[fill=blue!15, draw=blue!80!black, thick] (0.5,0.5) circle (0.15);
                \filldraw[fill=red!30, draw=red!80!black, thick] (0.5,0.5) circle (0.05);

                \draw[<->, >=stealth, densely dashed] (0.5,0.5) -- (0.5, 0.65) node[midway, left] {\tiny $r=0.15$};
                \draw[<->, >=stealth] (0.5,0.5) -- (0.55, 0.5) node[right] {\tiny $r=0.05$};
                \draw[->, >=stealth, red!80!black, thick] (0.5, 0.28) -- (0.5, 0.48);
                \node[red!80!black, below] at (0.5, 0.28) {\tiny Inclusion};
            \end{scope}
        \end{scope}

        \begin{scope}[xshift=2.8cm]
            \node[align=center] at (0.5, 1.1) {\small (c) Composite \& Overlap};
            \draw[thick, darkgray] (0,0) rectangle (1,1);

            \filldraw[fill=blue!10, draw=blue!80!black, dashed, even odd rule]
                (0.1,0.1) rectangle (0.9,0.9)
                (0.5,0.5) circle (0.1);

            \filldraw[fill=blue!30, draw=green!50!black, thick, fill opacity=0.4] (0.5,0.5) circle (0.15);
            \filldraw[fill=red!40, draw=red!80!black, thick] (0.5,0.5) circle (0.05);

            \fill[orange, opacity=0.5, even odd rule]
                (0.5,0.5) circle (0.15)
                (0.5,0.5) circle (0.10);

            \node[align=center, font=\linespread{0.8}\selectfont, orange!80!black] at (0.5, 0.22) {\scriptsize Overlap Zone\\ \scriptsize ($0.1 \le r \le 0.15$)};
            \draw[->, >=stealth, thick, orange!80!black] (0.56, 0.38) -- (0.56, 0.2);
        \end{scope}
    \end{tikzpicture}
    \caption{Setup for the Elastic Inclusion Problem. Overlap zone is an annular ring bridging $\Omega_B$ and $\Omega_S$.}
\end{figure}%
}

\newcommand{\figHertzContact}{%
\begin{figure}[htbp]
    \centering
    \begin{tikzpicture}[scale=1]

        \begin{scope}[xshift=0cm]
            \node[align=center] at (2.0, 3.2) {\small (a) Coarse Domain $\Omega_B$};

            \draw[step=0.2, gray!30, thin] (0, 0.5) grid (4.0, 2.5);

            \foreach \x in {1.0, 1.5, 2.0, 2.5, 3.0} {
                \draw[->, >=stealth, thick, red] (\x, 2.6) -- (\x, 2.1);
            }
            \node[red] at (2.0, 2.8) {\scriptsize Distributed Load};

            \begin{scope}
                \clip (0, 0.8) rectangle (4, 2.0);
                \filldraw[fill=blue!15, draw=blue!80!black, thick] (2,2) ++(180:1.6) arc (180:360:1.6) -- cycle;
            \end{scope}

            \draw[blue!80!black, thick] (0.4, 2.0) -- (3.6, 2.0); 
            \draw[blue!80!black, thick, densely dashed] (0.93, 0.8) -- (3.07, 0.8); 

            \node[blue!80!black] at (2.0, 1.5) {Bulk Body};
            \draw[<->, >=stealth, densely dashed] (4.2, 0.8) -- (4.2, 2.0) node[midway, right] {\scriptsize $y \ge 0.08$};
        \end{scope}

        \begin{scope}[xshift=5cm]
            \node[align=center] at (2.0, 3.2) {\small (b) Fine Domain $\Omega_S$};

            \draw[step=0.08, gray!50, thin] (0.5, 0.3) grid (3.5, 1.3);

            \begin{scope}
                \clip (0, 0) rectangle (4, 1.2);
                \filldraw[fill=green!20, draw=green!60!black, thick] (2,2) ++(180:1.6) arc (180:360:1.6) -- cycle;
            \end{scope}

            \draw[green!60!black, thick, densely dashed] (0.6, 1.2) -- (3.4, 1.2); 

            \node[green!40!black] at (2.0, 0.75) {\scriptsize Contact Pad};
            \draw[<->, >=stealth, densely dashed] (3.7, 0.4) -- (3.7, 1.2) node[midway, right] {\scriptsize $y \le 0.12$};
        \end{scope}

        \begin{scope}[xshift=10cm]
            \node[align=center] at (2.0, 3.2) {\small (c) Contact Setup};

            \draw[ultra thick, black] (0.5, 0.38) -- (3.5, 0.38);
            \fill[pattern=north east lines] (0.5, 0.18) rectangle (3.5, 0.38);
            \node[below] at (2.0, 0.18) {\scriptsize Rigid Boundary ($y=0$)};

            \begin{scope}
                \clip (0, 0) rectangle (4, 2.0);
                \filldraw[fill=blue!10, draw=blue!80!black, dashed, fill opacity=0.7] (2,2) ++(180:1.6) arc (180:360:1.6) -- cycle;

                \begin{scope}
                    \clip (0, 0) rectangle (4, 1.2);
                    \filldraw[fill=green!30, draw=green!60!black, thick, fill opacity=0.9] (2,2) ++(180:1.6) arc (180:360:1.6) -- cycle;
                \end{scope}
            \end{scope}
            \draw[blue!80!black, thick] (0.4, 2.0) -- (3.6, 2.0); 

            \begin{scope}
                \clip (0, 0.8) rectangle (4, 1.2);
                \fill[orange, opacity=0.7] (2,2) ++(180:1.6) arc (180:360:1.6) -- cycle;
            \end{scope}

            \draw[->, >=stealth, thick, orange!90!black] (3.0, 1.6) -- (3.0, 1.0);
            \node[left, orange!90!black, align=right] at (3.1, 1.6) {\scriptsize Overlap Zone };

            \draw[->, >=stealth, thick, red] (1.0, 2.6) -- (1.0, 2.1);
            \draw[->, >=stealth, thick, red] (3.0, 2.6) -- (3.0, 2.1);
        \end{scope}

    \end{tikzpicture}
    \caption{Setup of the Hertzian contact benchmark. The fine sub-domain ($\Omega_S$) resolves the anticipated contact zone near the rigid plane, whereas the coarse sub-domain ($\Omega_B$) represents the far-field bulk response through an overlapping transition region.}
\end{figure}%
}

\newcommand{\figCantileverSetup}{%
\begin{figure}[htbp]
    \centering
    \vspace{-0.15cm}
    \makebox[\textwidth][c]{\small Domain Layout}
    \vspace{0.12cm}
    \begin{tikzpicture}[scale=1.2]
        \draw[dashed, gray!30] (0, -0.2) rectangle (10, 1.2);

        \filldraw[fill=blue!10, draw=blue!80!black, dashed] (0.1, 0.4) rectangle (4.5, 0.8);
        \filldraw[fill=green!20, draw=green!60!black, dashed] (3.9, 0.4) rectangle (9.4, 0.8);

        \fill[orange, opacity=0.7] (3.9, 0.4) rectangle (4.5, 0.8);

        \draw[very thick] (0.1, 0.2) -- (0.1, 1.0);
        \foreach \y in {0.2, 0.4, 0.6, 0.8, 1.0} {
            \draw (0.1, \y) -- (-0.1, \y-0.2);
        }

        \draw[<->, >=stealth, blue!80!black, thick] (0.1, 1.12) -- (4.5, 1.12)
            node[midway, above] {\scriptsize $\Omega_L: [0.05, 0.45],\ \ell=0.40$};
        \draw[<->, >=stealth, green!60!black, thick] (3.9, 1.00) -- (9.4, 1.00)
            node[midway, above] {\scriptsize $\Omega_R: [0.39, 0.90],\ \ell=0.51$};

        \draw[<->, >=stealth, gray!70!black, thick] (9.65, 0.4) -- (9.65, 0.8)
            node[midway, right] {\scriptsize $t=0.02$};

        \draw[<->, >=stealth, orange!90!black, thick] (3.9, 0.2) -- (4.5, 0.2) node[midway, below] {\scriptsize Overlap ($x \in [0.39, 0.45]$)};
    \end{tikzpicture}
    \caption{Spatial domain decomposition for the cantilever beam. The structure is dynamically coupled via the intermediate orange overlap zone. The blue domain has width $1.0$ and resolution $200 \times 200$, while the green domain has width $0.5$ and resolution $100 \times 100$.}
    \label{fig:cantilever_setup}
\end{figure}%
}

%% file: sections/introduction.tex

\section{Introduction}
\label{sec:intro}
The material point method (MPM) has become an important computational framework in continuum mechanics \cite{bardenhagen2004generalized, de2020material}. It combines a Lagrangian description of the material through moving particles that carry the state variables with an Eulerian background grid used to solve the governing equations \cite{jiang2016material,wikeckowski2004material}. This hybrid particle–grid structure makes MPM particularly attractive for problems involving large deformations \cite{wikeckowski2004material}, history-dependent constitutive behaviour \cite{solowski2015evaluation,sulsky1994particle,baumgarten2019general,li2022energetically}, evolving interfaces \cite{jiang2020hybrid}, and self-contact \cite{jiang2016material, xiao2021dp}, where conventional mesh-based approaches such as the finite element method may suffer from severe mesh distortion. Owing to these advantages, MPM has been successfully applied to a wide range of problems in computational mechanics, including geophysical flows \cite{jiang2023erosion,jiang2022hybrid,liang2023revealing,zhao2023coupled}, impact \cite{jiang2024impact,ma2009comparison} and penetration \cite{huang2011contact,jiang2022hybrid}, and fracture \cite{hu2023explicit} and fragmentation of solids \cite{xiao2021dp}. 

Despite these advantages, MPM may still become computationally demanding when high accuracy is required \cite{bardenhagen2000material,sulsky1994particle}. At each computational step, information must be transferred repeatedly between the material points and the background grid, and this cost increases rapidly as the particle density and grid resolution are refined \cite{de2020material,jiang2016material}. In addition, the standard MPM commonly evaluates the weak-form integrals by treating the moving material points as quadrature points, which may introduce spatial integration errors that are difficult to control \cite{steffen2008analysis,wilson2021distillation}. These errors can degrade the accuracy of computed stress and strain fields, especially in problems with strongly localized deformation or high solution gradients \cite{zhang2011material}. Another practical difficulty is that the evolving material boundary is typically not represented by conforming grid segments or boundary nodes, which can make accurate boundary description and boundary-condition imposition more difficult in standard MPM formulations \cite{bardenhagen2000material,cortis2018imposition,singer2024lagrange}. This further increases the demand for high local resolution near curved, moving, or strongly deforming boundaries. Consequently, sufficiently accurate MPM simulations often require both a dense particle discretization and a fine background grid, which substantially increases the overall computational expense. Improving computational efficiency while retaining local accuracy has therefore become a central issue in the further development of MPM \cite{lian2014tied,lian2015mesh,ma2006structured}.

However, local mesh refinement in MPM is algorithmically more challenging than in conventional mesh-based methods, because MPM decouples the material description with the computational grids \cite{sulsky1994particle}. As a result, local refinement does not only modify the computational grid, but also affects the particle--grid transfer operators and the consistency of the associated interpolation \cite{lian2015mesh,tan2002hierarchical}. In many improved MPM formulations, higher-order basis functions, with support spanning multiple background cells, are introduced to improve the smoothness of the deformation-gradient field and to mitigate cell-crossing-related numerical artefacts \cite{bardenhagen2004generalized,gan2018enhancement,moutsanidis2020iga}. Under local refinement, this wider support makes the transition across coarse—fine interfaces considerably more delicate. Because fundamental properties of the kernel or shape functions – such as partition of unity, non-negativity, local support, and polynomial reproduction – must be preserved across nonuniform resolution levels and cannot be maintained automatically \cite{sadeghirad2011convected,zhang2011material,zhang2021truncated}. Consequently, constructing accurate and stable kernel functions near coarse–fine interfaces becomes a central difficulty in locally refined MPM formulations. In addition, when local refinement is implemented on structured Cartesian grids, the treatment of hanging nodes introduces further complications \cite{lian2015mesh,lian2014tied,sun2020local}. Although unstructured background grids provide greater geometric flexibility, they are less straightforward in this context because the construction of sufficiently smooth basis functions is more intricate, and their data structures are generally less favorable for the implementation of parallel computation \cite{cao2025unstructured,de2021extension}.

Regardless of these challenges, substantial progress has been made in the development of local-refinement strategies for MPM. Broadly speaking, existing approaches may be divided into two main categories. One category seeks to realize local refinement through modifications of the kernel functions themselves \cite{gao2017Adaptive,lian2015mesh,lian2014tied,ma2006structured,ma2005multiscale,tan2002hierarchical,woo2018simulation}. Early developments explored refinement schemes for standard MPM with integer refinement ratios, which were later extended to GIMP-type formulations in order to improve smoothness across refinement transitions \cite{ma2006structured,ma2005multiscale,tan2002hierarchical}. Subsequent studies further increased the flexibility of these approaches by introducing special treatments of the kernel functions in coarse-fine transition zones together with compatibility constraints associated with hanging nodes \cite{barzoki2025novel,lian2015mesh,lian2014tied,woo2018simulation}. Along a different line, higher-order kernel functions such as B-splines have been combined with hierarchical and truncated hierarchical constructions, so that local refinement can be introduced while preserving desirable properties such as enhanced smoothness and improved approximation quality \cite{zhang2021truncated}. A second category of methods attempts to retain the original kernel functions within the coarse and fine subdomains and instead addresses the refinement transition through special interface-coupling or bridging treatments \cite{he2025multi,sun2020local}. In such approaches, displacement compatibility across the coarse-fine interface is typically enforced by means of Lagrange multipliers or penalty constraints.

Although these developments have substantially advanced local refinement in MPM, their effectiveness is often accompanied by non-negligible algorithmic overhead. In the kernel modification category, local refinement is typically achieved through specially constructed transition functions, hierarchical data structures \cite{gao2017Adaptive,zhang2021truncated}, or coarse-fine treatments that are not part of the standard MPM formulation \cite{de2021extension,lian2015mesh,lian2014tied}. While effective, such constructions may considerably increase implementation complexity and become less attractive in problems involving large deformation, strong particle migration, or highly dynamic kinematics, where particles interact repeatedly with refinement-transition regions. The associated issues may become even more cumbersome when multiple refined subdomains are present, as in laminated composites or particle-reinforced materials. In this sense, the added refinement machinery may offset part of the simplicity and modularity that make the standard MPM attractive in the first place. By contrast, interface-constraint approaches retain the original kernel functions within the coarse and fine subdomains, but they introduce a different class of difficulties. Lagrange-multiplier-based formulations lead to saddle-point systems whose stability depends on a proper choice of discrete spaces and which may complicate the algebraic structure through additional interface unknowns \cite{sun2020local}. Penalty-based formulations avoid extra multiplier fields, but their accuracy and stability depend sensitively on the penalty parameter: an insufficient penalty may result in noticeable interface mismatch, whereas an excessively large penalty may cause artificial stiffness and severe ill-conditioning \cite{he2025multi}. This motivates the development for an alternative formulation in which local refinement can retain standard subdomain MPM discretization while shifting the coupling complexity to a more modular and mathematically analyzable interface iteration.

In this work, we develop an overlapping Schwarz framework \cite{schwarz1870ueber,mota2017schwarz} for the material point method, namely OS-MPM, that enables concurrent coarse–fine coupling over subdomains with heterogeneous spatial and temporal resolutions. The proposed formulation realizes local space–time refinement through an MPM-specific interface treatment consisting of mass-weighted spatial transmission, temporal interpolation, fine-scale subcycling, and iteration-consistent restoration of particle states. In contrast to local-refinement strategies that rely on globally modified basis functions or strongly enforced interface constraints, the present approach retains standard subdomain MPM discretizations and handles the coarse–fine interaction through non-matching-grid transmission operators within the Schwarz alternating procedure. In this way, OS-MPM provides a practical and modular route to local h-refinement and local temporal subcycling in MPM. Numerical results show that the proposed framework delivers substantial computational savings while preserving the accuracy and convergence behavior of monolithic fine-resolution simulations.







%% file: sections/formulation.tex
\section{Governing Framework}\label{sec:formulation}
In this section, we formulate the domain decomposed problem addressed in this work, summarize the continuum mechanics and MPM discretization framework, and introduce the overlapping Schwarz method that forms the algorithmic basis of our approach.

\subsection{Problem Statement}\label{subsec:problem_statement}
Consider a material domain $\Omega_0 \subset \mathbb{R}^d$ that contains localized regions of high-frequency deformation, such as impact zones and shear bands. These regions require fine spatial and temporal resolution to accurately capture the underlying physics, while the bulk of the domain deforms on a significantly coarser scale. Under implicit time integration, the state at each time step is obtained by solving an energy minimization problem:
\begin{equation}
    x_{n+1} = \arg \min_{x \in \mathcal{V}} E(x),
    \label{eq:global_minimization}
\end{equation}
where $E(x)$ is the incremental potential \cite{kane2000variational} that encompasses inertia, elasticity, and body force over $\Omega_0$ (see more details in Section~\ref{subsec:continuum_mpm}).

When the entire domain is discretized at the fine resolution required by the localized phenomena, with grid spacing $\Delta x_S$ and time step $\Delta t$, the computational cost becomes prohibitive. The size of the resulting nonlinear system scales as $\mathcal{O}(|\Omega_0|/\Delta x_S^d)$, and the total number of time steps scales as $\mathcal{O}(T/\Delta t)$, rendering a monolithic solve impractical for large-scale problems.

The key observation is that only a small subdomain requires such fine resolution. This motivates a domain decomposition strategy: $\Omega_0$ is partitioned into overlapping subdomains with heterogeneous spatial and temporal resolutions, and the global problem \eqref{eq:global_minimization} is solved iteratively by alternating between subdomain solves coupled through interface conditions. The mathematical foundation for this approach, the overlapping Schwarz alternating method, is introduced in Section~\ref{subsec:schwarz_background}, and we specialize it to the MPM setting in Section~\ref{sec:method}.

\subsection{Variational MPM Discretization}\label{subsec:continuum_mpm}
In this section, we start from the Eulerian form of the momentum equation, derive its weak form, and spatially discretize it on a structured grid. We then recast the implicit time integration into a variational energy minimization problem \cite{wang2020hierarchical}. Finally, we introduce the Material Point Method (MPM) \cite{sulsky1994particle} with APIC transfer \cite{jiang2017angular} to evaluate the associated spatial integrals and kinematics via particle quadrature.

\subsubsection{Strong and Weak Forms of the Governing Equation}
Let $\Omega^t$ be the current domain of the material at time $t$. The strong form of the momentum equation in Eulerian view is given by:
\begin{equation}
    \rho \frac{D\mathbf{v}}{Dt} = \nabla^{\mathbf{x}} \cdot \boldsymbol{\sigma} + \rho \mathbf{g},  \label{eq:strong_form}
\end{equation}
where $\rho$ is the current mass density, $\mathbf{v}$ is the velocity field, ${D\mathbf{v}}/{Dt}=\dot{\mathbf{v}}+\mathbf{v}\cdot\nabla \mathbf{v}$ denotes the material derivative of the velocity, $\boldsymbol{\sigma}$ is the Cauchy stress tensor, $\mathbf{g}$ represents acceleration of body forces such as gravity, and $\nabla^{\mathbf{x}}$ denotes the spatial gradient operator, where the superscript $\mathbf{x}$ indicates differentiation with respect to Eulerian (spatial) coordinates, as opposed to Lagrangian (material/reference) coordinates $\mathbf{X}$.

To derive the weak form, we multiply the strong form in Equation \eqref{eq:strong_form} by an arbitrary vector-valued test function $\mathbf{q}(\mathbf{x})$ that vanishes on the Dirichlet boundary, and integrate over the current domain $\Omega^t$. Applying integration by parts and the divergence theorem to the stress term yields the Eulerian weak form:
\begin{equation}
    \int_{\Omega^t} \rho \frac{D\mathbf{v}}{Dt} \cdot \mathbf{q} \, d\mathbf{x} + \int_{\Omega^t} \boldsymbol{\sigma} : \nabla^{\mathbf{x}} \mathbf{q} \, d\mathbf{x} = \int_{\Omega^t} \rho \mathbf{g} \cdot \mathbf{q} \, d\mathbf{x} + \int_{\partial \Omega_{\tau}^t} \boldsymbol{\tau} \cdot \mathbf{q} \, ds \label{eq:weak_form}
\end{equation}
where $\mathbf{n}$ is the outward unit normal, and $\boldsymbol{\tau} = \boldsymbol{\sigma}\mathbf{n}$ is the boundary traction.

\subsubsection{Eulerian Grid Discretization}
To spatially discretize the continuum domain, we first introduce a structured Eulerian background grid equipped with nodal shape functions $N_i(x)$. The grid defines a finite-dimensional interpolation space through which continuous test functions and material fields are projected onto their discrete nodal representations for the weak-form discretization of the linear momentum balance in Eq.\eqref{eq:strong_form}. Specifically, both the arbitrary test function $\mathbf{q}$ and material fields $\boldsymbol{\varphi}$ are approximated using these interpolants as 
\begin{equation}
\mathbf{q}(\mathbf{x}) = \sum_i \mathbf{q}_i N_i(\mathbf{x}) \;, \; \boldsymbol{\varphi}(\mathbf{x}) = \sum_i  N_i(\mathbf{x})\boldsymbol{\varphi}_i.
\label{eq:interpolation}
\end{equation}

With the above Eulerian interpolation, the weak form is projected onto the grid-based test space to obtain the semi-discrete nodal momentum balance. The convective contribution contained in the material derivative is not assembled as an independent Eulerian advection operator at this stage; in the MPM formulation, material transport is introduced later through the material update described in Section~2.2.4. Substituting the interpolated test function and material fields into Eq.~\eqref{eq:weak_form}, and collecting the coefficients of the arbitrary nodal variations $\mathbf{q}_i$ yields
\begin{equation}
    \sum_j m_{ij} \dot{\mathbf{v}}_j = \mathbf{f}_i^{\text{int}} + \mathbf{f}_i^{\text{ext}}, \label{eq:discrete_momentum}
\end{equation}
where the consistent mass matrix $m_{ij}$, the internal force $\mathbf{f}_i^{\text{int}}$, and the external force $\mathbf{f}_i^{\text{ext}}$ are defined by the spatial integrals over the current domain:
\begin{align}
    m_{ij} &= \int_{\Omega^t} \rho N_i(\mathbf{x}) N_j(\mathbf{x}) \, d\mathbf{x}, \label{eq:continuous_mass}\\
    \mathbf{f}_i^{\text{int}} &= - \int_{\Omega^t} \boldsymbol{\sigma} \cdot \nabla^{\mathbf{x}} N_i(\mathbf{x}) \, d\mathbf{x}, \label{eq:continuous_fint}\\
    \mathbf{f}_i^{\text{ext}} &= \int_{\Omega^t} \rho \mathbf{g} N_i(\mathbf{x}) \, d\mathbf{x} + \int_{\partial \Omega_{\tau}^t} \boldsymbol{\tau} N_i(\mathbf{x}) \, ds. \label{eq:continuous_fext}
\end{align}

\subsubsection{Implicit Time Integration and Variational Form}
For dynamic simulation and numerical stability, we apply implicit time integration (e.g., Backward Euler) to the semi-discrete Equation \eqref{eq:discrete_momentum}. Assuming a standard mass lumping approximation ($m_{i}  = \sum_j m_{ij}$), the time-discretized residual equation for the unknown grid velocity $\mathbf{v}^{n+1}$ is formulated as:
\begin{equation}
    \mathbf{r}_i(\mathbf{v}^{n+1}) = m_i \frac{\mathbf{v}_i^{n+1} - \mathbf{v}_i^n}{\Delta t} - \mathbf{f}_i^{\text{int}}(\mathbf{v}^{n+1}) - \mathbf{f}_i^{\text{ext}} = \mathbf{0}.
    \label{eq:grid_residual}
\end{equation}
To express the backward-Euler update in terms of displacement-like nodal unknowns, we introduce the trial nodal position $\hat{\mathbf{x}}_i = \mathbf{x}_i^n + \Delta t \mathbf{v}_i^{n+1}$ to parameterize the velocity degrees of freedom, recasting the residual as $\mathbf{r}(\hat{\mathbf{x}})$. 

For constitutive models admitting a variational potential, the internal force is assumed to be derivable as the negative derivative of the corresponding stored or incremental energy. In the linear and hyper-elastic setting considered in the present examples, with strain-energy density $\Psi$, the elastic potential and the corresponding force are written as
\begin{equation}
\Phi(\hat{\mathbf{x}}) = \int_{\Omega^0} \Psi(\mathbf{F}(\hat{\mathbf{x}})) \, d\mathbf{X}\;,\;\mathbf{f}_i^{\text{int}}(\hat{\mathbf{x}}) = -\frac{\partial \Phi}{\partial \hat{\mathbf{x}}_i}.
\end{equation}
By analytically integrating the momentum residual Eq.~\eqref{eq:grid_residual} over the spatial DOFs $\hat{\mathbf{x}}_i$, the implicit update can be rigorously cast as an unconstrained optimization problem:
\begin{equation}
    \hat{\mathbf{x}} = \underset{\hat{\mathbf{x}}}{\arg\min} \ E_h(\hat{\mathbf{x}})
\end{equation}
where the total incremental potential \cite{kane2000variational} is defined as:
\begin{equation}
    E_h(\hat{\mathbf{x}}) = \sum_{i} \frac{m_i}{2\Delta t^2} \|\hat{\mathbf{x}}_i - \tilde{\mathbf{x}}_i\|^2 + \Phi(\hat{\mathbf{x}}) - \sum_i \hat{\mathbf{x}}_i \cdot \mathbf{f}_{i}^{\text{ext}},
    \label{eq:discrete_energy}
\end{equation}
with $\tilde{\mathbf{x}}_i = \mathbf{x}_i^n + \Delta t \mathbf{v}_i^n$. Taking the gradient of $E_h(\hat{\mathbf{x}})$ with respect to $\hat{\mathbf{x}}_i$ and setting it to zero recovers the original implicit momentum equation. This variational perspective enables using modern optimization techniques, such as line search, to guarantee global convergence, leading to robust implicit time integration.

\subsubsection{Material Point Discretization }\label{subsubsec:mpm_particles}
At this stage, the formulation still contains continuous spatial integrals in the nodal mass and force terms. We employ the Material Point Method (MPM) \cite{sulsky1994particle} to evaluate these integrals by material-point quadrature, with material transport accounted for through the advection of material points. The material state is tracked by Lagrangian particles (indexed by $p$) with mass $m_p$, position $\mathbf{x}_p$, velocity $\mathbf{v}_p$, and deformation gradient $\mathbf{F}_p$, satisfying mass conservation by construction.

At the beginning of each time step $n$, particle mass and momentum are transferred to the grid (i.e., P2G) to initialize the nodal mass $m_i$ and velocity $\mathbf{v}_i^n$. We utilize the Affine Particle-in-Cell (APIC) scheme \cite{jiang2015apic} to preserve angular momentum:
\begin{equation}
    m_i = \sum_{p} m_p N_i(\mathbf{x}_p^n), \quad (m\mathbf{v})_i^n = \sum_{p} m_p N_i(\mathbf{x}_p^n) \left( \mathbf{v}_p^n + \mathbf{B}_p^n (\mathbf{D}_p^n)^{-1} (\mathbf{x}_i - \mathbf{x}_p^n) \right),
    \label{eq:p2g_transfer}
\end{equation}
where $\mathbf{B}_p^n$ is the affine velocity matrix and $\mathbf{D}_p^n$ is the particle inertia tensor. This transfer provides the initial nodal velocity $\mathbf{v}_i^n$, from which the inertial target position $\tilde{\mathbf{x}}_i = \mathbf{x}_i^n + \Delta t \mathbf{v}_i^n$ is computed for the variational update in Eq.~\eqref{eq:discrete_energy}.

Furthermore, MPM utilizes the particles as quadrature points to evaluate the continuous integrals. Applying particle quadrature to the continuous force definitions (Eqs. \ref{eq:continuous_fint}-\ref{eq:continuous_fext}), the external and internal forces are evaluated as:
\begin{equation}
    \mathbf{f}_i^{\text{ext}} \approx \sum_p m_p \mathbf{g} N_i(\mathbf{x}_p^n) + \mathbf{f}_{i}^{\text{trac}},
\end{equation}
where the first term is the particle quadrature approximation of the body force integral, and $\mathbf{f}_i^{\text{trac}} = \int_{\partial\Omega_\tau^t} N_i(\mathbf{x})\,\boldsymbol{\tau}\,ds$ is the nodal traction force arising from the Neumann boundary condition $\boldsymbol{\tau} = \boldsymbol{\sigma}\mathbf{n}$ on $\partial\Omega_\tau^t$, and

\begin{equation}
    \mathbf{f}_i^{\text{int}} \approx - \sum_p V_p^t \boldsymbol{\sigma}_p \cdot \nabla^{\mathbf{x}} N_i(\mathbf{x}_p^n) = - \sum_p V_p^0 (\mathbf{P}_p \mathbf{F}_p^T) \cdot \nabla^{\mathbf{x}} N_i(\mathbf{x}_p^n).
\end{equation}

Crucially, to evaluate the total elastic potential $\Phi(\hat{\mathbf{x}})$ and its gradient, we must relate the virtual grid deformation $\hat{\mathbf{x}}$ to the local material state. This is achieved via the updated Lagrangian formulation for tracking deformation gradient. Using the gradient of the grid shape functions, $\mathbf{F}_p$ is updated purely as a function of the virtual grid positions when backward Euler is used for Eq.~\eqref{eq:discrete_momentum}:
\begin{equation}
    \mathbf{F}_p(\hat{\mathbf{x}}) \approx \mathbf{F}_p^n + \Delta t \dot{\mathbf{F}}_p^{n+1} = \left( \mathbf{I} + \sum_i (\hat{\mathbf{x}}_i - \mathbf{x}_i^n) \left(\nabla^{\mathbf{x}} N_i(\mathbf{x}_p^n)\right)^T \right) \mathbf{F}_p^n.
    \label{eq:F_update}
\end{equation}
Consequently, the continuous elastic potential is discretized into a computable sum over particles, $\Phi(\hat{\mathbf{x}}) \approx \sum_p V_p^0 \Psi(\mathbf{F}_p(\hat{\mathbf{x}}))$, which completely defines the discrete objective function $E_h(\hat{\mathbf{x}})$ in a particle-based discretization.

Once grid update detailed in the next subsection is completed, the particle states are updated using the virtually displaced grid position $\hat{\mathbf{x}}$ and velocity $\mathbf{v}_i^{n+1}$.
The velocity and affine matrix of each particle are updated via the APIC grid-to-particle transfer (i.e., G2P):
\begin{align}
    \mathbf{v}_p^{n+1} &= \sum_i N_i(\mathbf{x}_p^n)\, \mathbf{v}_i^{n+1}, \label{eq:g2p_v} \\
    \mathbf{B}_p^{n+1} &= \sum_i N_i(\mathbf{x}_p^n)\, \mathbf{v}_i^{n+1} (\mathbf{x}_i - \mathbf{x}_p^n)^T. \label{eq:g2p_B}
\end{align}
The deformation gradient $\mathbf{F}_p^{n+1}$ is evaluated from $\hat{\mathbf{x}}$ via Eq.~\eqref{eq:F_update}.
Finally, particles are advected to their new positions to account for the $\mathbf{v}\cdot\nabla \mathbf{v}$ term:
\begin{equation}
    \mathbf{x}_p^{n+1} = \mathbf{x}_p^n + \Delta t\, \mathbf{v}_p^{n+1}. \label{eq:advection}
\end{equation}

\subsubsection{Nonlinear Minimization via Newton's Method with Line Search}
To solve for $\hat{\mathbf{x}}$, we minimize the nonlinear objective defined in Eq. \eqref{eq:discrete_energy} using Newton’s method. At each iteration $k$, we solve the linear system for the spatial increment $\delta \hat{\mathbf{x}}$:
\begin{equation}
    \left( \frac{\mathbf{M}}{\Delta t^2} + \mathbf{K}(\hat{\mathbf{x}}^{(k)}) \right) \delta \hat{\mathbf{x}} = -\mathbf{r}(\hat{\mathbf{x}}^{(k)}),
\end{equation}
where $\mathbf{K} = \nabla^2_{\hat{\mathbf{x}}} \Phi$ is the tangent stiffness matrix (the Hessian of the elastic potential), and $\mathbf{M}$ is the diagonal lumped mass matrix. To ensure global convergence and a monotonic decrease of the energy functional $E_h$, the Newton increment is coupled with backtracking line search \cite{nocedal2006numerical}, which searches for a sufficiently small step size $\alpha^{(k)} \in (0, 1]$ starting from $1$ by iteratively halving it until $E_h(\hat{\mathbf{x}}^{(k+1)}) <E_h(\hat{\mathbf{x}}^{(k)})$ is satisfied at the new iterate $\hat{\mathbf{x}}^{(k+1)} = \hat{\mathbf{x}}^{(k)} + \alpha^{(k)} \delta \hat{\mathbf{x}}$.

\subsubsection{Progressive Solution for Static Equilibrium}
While the formulation described above is dynamic in nature, it is highly effective and frequently employed to solve for quasi-static or static equilibrium in computational mechanics \cite{guilkey2003implicit, beuth2012formulation}. In Lagrangian mesh-based methods, static equilibrium can often be obtained via a single nonlinear solve over the entire load step. However, this single-step static approach is unreliable under
extreme deformation due to mesh distortion; a Total-Lagrangian MPM (TL-MPM) formulation \cite{charlton2017fully} faces an analogous difficulty, as it relies on fixed particle-grid associativity defined in the initial configuration, which prevents it from naturally resolving large geometric changes and contact.

Therefore, in our framework, we progressively solve for the static equilibrium by advancing the system through a sequence of backward Euler time steps. By taking sufficiently large time steps under the CFL condition, the inertial coefficient $m_i/\Delta t^2$ in $E_h$ becomes small relative to the elastic terms, reducing each time step to an effective quasi-static solve; due to the L-stability of Backward Euler, it suppresses any residual transient oscillations \cite{guilkey2003implicit}, allowing the system to robustly converge to the static equilibrium state while naturally updating the particle-grid associativity at each time step.

\subsection{Overlapping Schwarz Alternating Method}\label{subsec:schwarz_background}
The Schwarz alternating method is a classical iterative technique for solving boundary value problems on decomposed domains, enabling multi-resolution coupling. Originally introduced by Schwarz \cite{schwarz1870ueber} to prove the existence of harmonic functions on irregular domains, the method was placed on rigorous mathematical footing by Lions \cite{lions1988schwarz,lions1990schwarz}, who established convergence for both overlapping and non-overlapping variants. In addition to directly serving as algebraic system solvers, using domain decomposition to design preconditioners has also been explored \cite{guo2024barrier,huang2025stiffgipc,li2019decomposed}. The comprehensive treatment of domain decomposition algorithms and their convergence theory can be found in Toselli and Widlund \cite{toselli2004domain}.

\begin{figure}[htbp]
    \centering
    \includegraphics[width=0.4\textwidth]{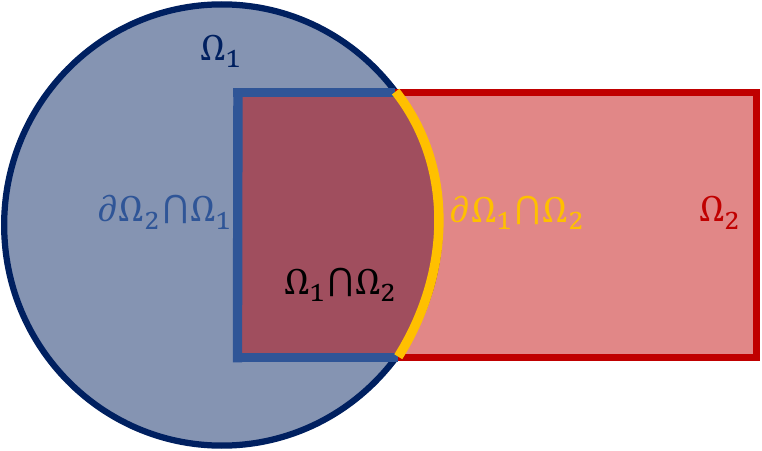}
    \caption{Illustration of overlapping subdomains.}
    \label{fig:overlap_subdomains}
\end{figure}

Consider a domain $\Omega$ decomposed into two overlapping subdomains $\Omega_1$ and $\Omega_2$ with non-empty overlap region $\Omega_{ovlp} = \Omega_1 \cap \Omega_2$ (see Fig.~\ref{fig:overlap_subdomains}). For a boundary value problem $\mathcal{L}u = f$ in $\Omega$, the multiplicative Schwarz alternating method generates iterates $\{u^{(k)}\}$ by alternately solving Dirichlet subproblems: at iteration $k+1$, one solves
\begin{align}
    \mathcal{L} u_1^{(k+1)} &= f \text{ in } \Omega_1, \quad u_1^{(k+1)} = u_2^{(k)} \text{ on } \partial\Omega_1 \cap \Omega_2, \label{eq:schwarz_sub1}\\
    \mathcal{L} u_2^{(k+1)} &= f \text{ in } \Omega_2, \quad u_2^{(k+1)} = u_1^{(k+1)} \text{ on } \partial\Omega_2 \cap \Omega_1. \label{eq:schwarz_sub2}
\end{align}

For problems admitting a variational formulation, such as the energy minimization in Eq.~\eqref{eq:global_minimization} arising from implicit time integration, this procedure is equivalent to an alternating minimization (block-coordinate descent), where each subdomain solve decreases the global energy functional restricted to the subdomain degrees of freedom while holding the complementary subdomain fixed.

It is well established that for elliptic and coercive operators, the Schwarz iteration is a contraction mapping with a geometric convergence rate $\rho < 1$ that depends on the overlap width $\delta$ \cite{lions1988schwarz,toselli2004domain}. Specifically, the error contracts as $\|e^{(k+1)}\| \le \rho \|e^{(k)}\|$, with $\rho$ decreasing (i.e., faster convergence) as the overlap $\delta$ increases. 
For nonlinear variational problems, overlapping Schwarz iterations have also been shown to converge geometrically under suitable well-posedness assumptions on the subdomain problems, including coercivity and quasi-convexity of the energy density. This result provides the theoretical foundation for applying the Schwarz alternating procedure to the finite-deformation setting considered in this work \cite{mota2017schwarz}.

The variant employed in this work is the multiplicative Schwarz method with Dirichlet-Dirichlet transmission conditions: each subdomain solve is a complete boundary value problem with Dirichlet data imposed on the artificial interface. In Section~\ref{sec:method}, we specialize this framework to the MPM setting, addressing the unique challenges of asynchronous time integration and non-conforming grids.

%% file: sections/method.tex
\section{Method: Overlapping Schwarz MPM}
\label{sec:method}

Having introduced the overlapping Schwarz framework in Section~\ref{subsec:schwarz_background}, we now specialize it to the MPM setting. The monolithic minimization of the incremental potential $E_h(\hat{\mathbf{x}})$ is replaced by the iterative Schwarz procedure applied to two subdomains with heterogeneous spatial and temporal resolutions.
This section details the concrete realization of the Schwarz iteration for MPM, including the spatial partitioning, mass-weighted transmission operators, and the sub-cycled fine subdomain evolution.

\subsection{Domain Partitioning}
We decompose the material domain $\Omega_0$ into two overlapping subdomains $\Omega_B$ and $\Omega_S$, such that $\Omega_B \cup \Omega_S = \Omega_0$. The intersection is denoted as $\Omega_{ovlp} = \Omega_B \cap \Omega_S \neq \emptyset$. The domains are discretized with heterogeneous resolutions:
\begin{itemize}
    \item \textbf{Background Domain ($\Omega_B$):} Characterized by coarse grid spacing $\Delta x_B$ and a large time step $\Delta T$.
    \item \textbf{Refined Domain ($\Omega_S$):} Characterized by fine grid spacing $\Delta x_S$ and a small time step $\Delta t$, where $\Delta x_S < \Delta x_B$ and $\Delta t \le \Delta T$.

\end{itemize}

Each domain carries its own independent set of Lagrangian particles: $\mathcal{P}_B$ for $\Omega_B$ and $\mathcal{P}_S$ for $\Omega_S$. Particles in the overlap region $\Omega_{ovlp}$ belong to both sets; their coupling is enforced through grid boundary conditions rather than direct particle interaction. For simplicity, we define the \textbf{sub-cycling ratio} $M = \Delta T / \Delta t$, assuming that $M \ge 1$ is an integer. While $\Omega_B$ evolves in a single step from $t_n$ to $t_{n+1}$, $\Omega_S$ performs $M$ sub-steps.

\subsection{Interface Operators}
To enforce kinematic compatibility in $\Omega_{ovlp}$, nodal velocities are exchanged between the non-conforming grids. We introduce projection and interpolation operators that help define Dirichlet boundary conditions for coupling the subdomain solvers.

\subsubsection{Boundary Grid Selection}
Before the simulation starts, each subdomain $\Omega_\alpha$, $\alpha \in \{B,S\}$, precomputes a set of boundary particles $\mathcal{P}_\alpha^\partial \subset \mathcal{P}_\alpha$ from the outer particle layer of that subdomain. These markers remain attached to the particles throughout the simulation. At the beginning of each global frame $t_n$, after the initial P2G transfer on each domain, these boundary particles are scattered to the corresponding background grid to determine the Dirichlet coupling nodes for the current frame:
\begin{equation}
    \Gamma_\alpha^n =
    \left\{ i \in \mathcal{N}_\alpha^n \;\middle|\;
    \sum_{p \in \mathcal{P}_\alpha^\partial}
    m_p N_{\alpha,i}(\mathbf{x}_p^n) > \tau_\Gamma
    \right\}, \qquad \alpha \in \{B,S\},
    \label{eq:boundary_grid_set}
\end{equation}
where $\mathcal{N}_\alpha^n$ is the active grid-node set of $\Omega_\alpha$ at time $t_n$, $N_{\alpha,i}$ is the grid shape function on that domain, and $\tau_\Gamma$ is a small mass threshold. Only nodes in $\Gamma_\alpha^n$ are eligible to receive Schwarz Dirichlet boundary conditions; all other active nodes are treated as ordinary free or interior grid nodes. The selected sets are kept fixed during the Schwarz iterations of the current global frame and are recomputed at the next frame after the new initial P2G transfer.

\begin{figure}[htbp]
    \centering
    \begin{subfigure}[t]{0.49\textwidth}
        \centering
        \includegraphics[width=\linewidth]{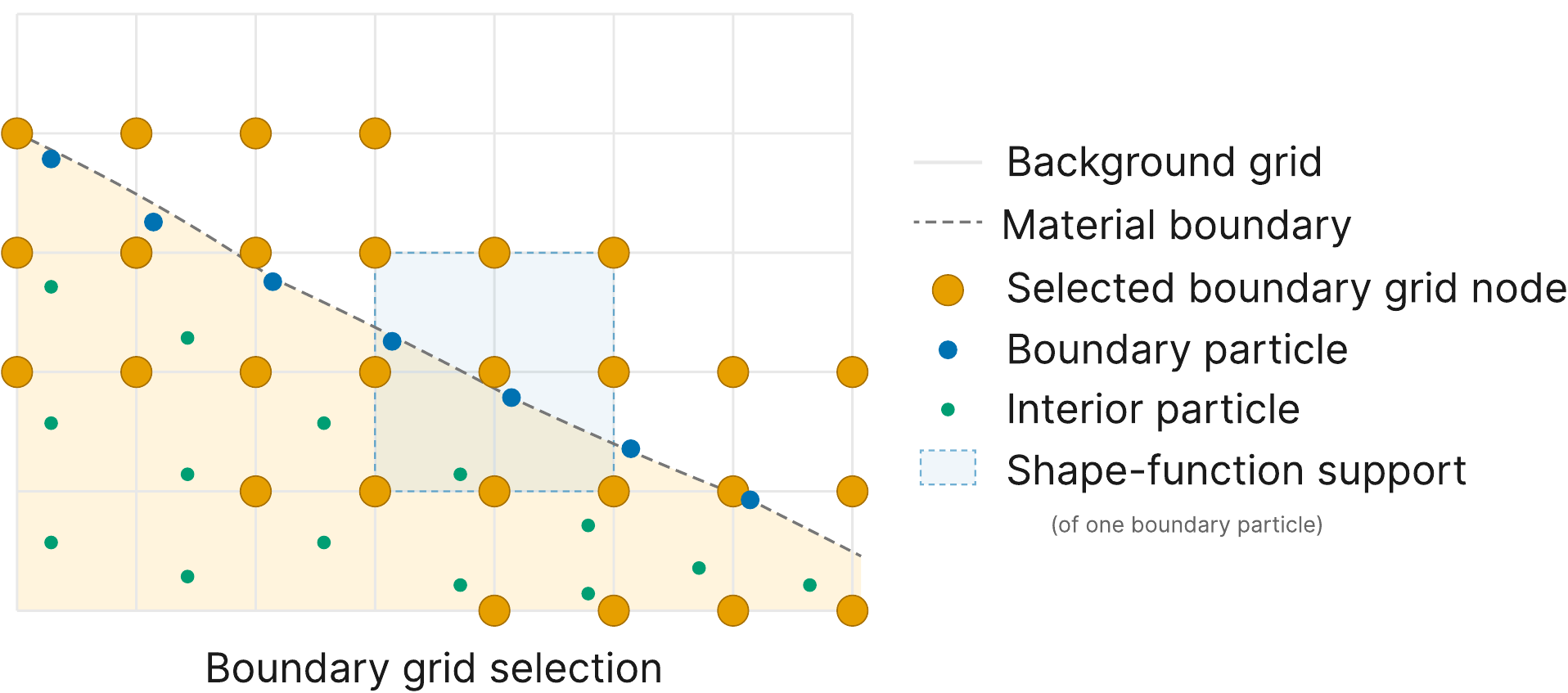}
        \caption{Boundary grid selection from boundary-particle P2G contributions.}
        \label{fig:boundary_grid_selection}
    \end{subfigure}\hfill
    \begin{subfigure}[t]{0.49\textwidth}
        \centering
        \includegraphics[width=\linewidth]{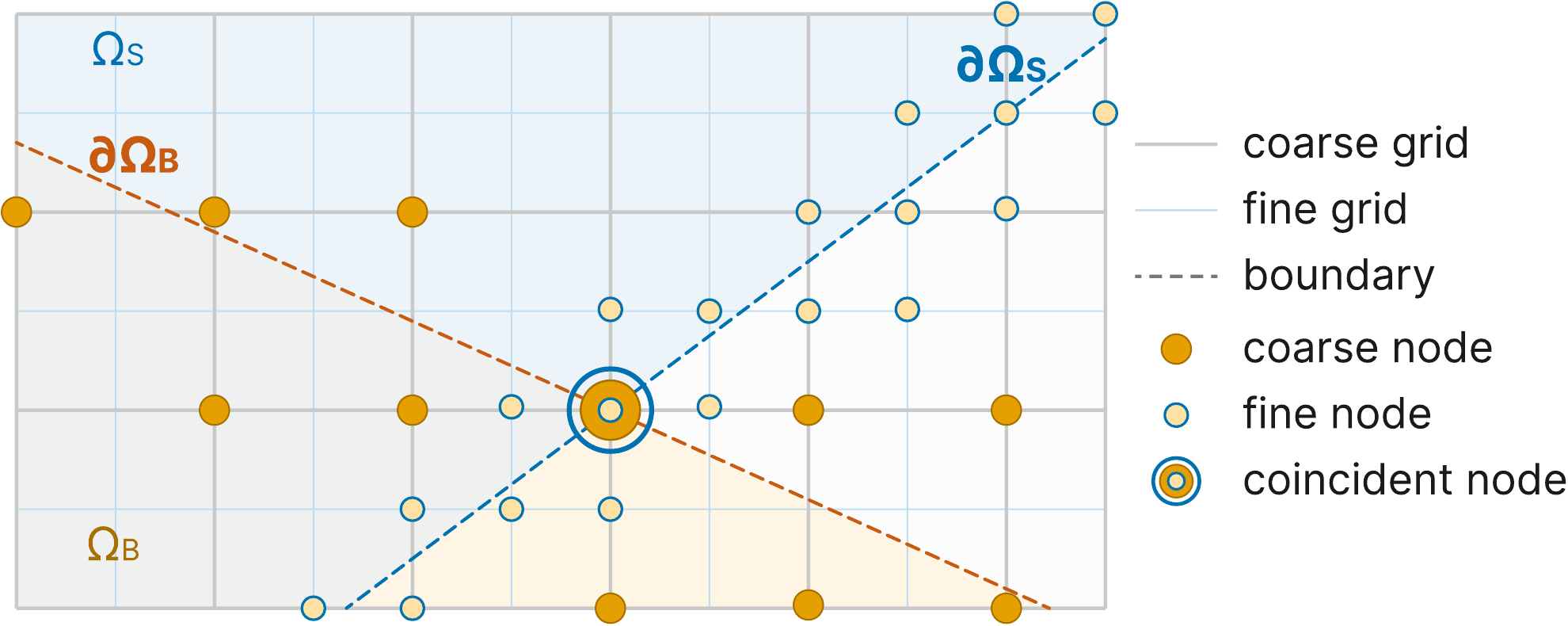}
        \caption{Boundary-node ambiguity caused by coincident coarse- and fine-grid boundary nodes.}
        \label{fig:boundary_node_ambiguity}
    \end{subfigure}
    \caption{Construction of boundary grid nodes for Schwarz coupling. The selected grid nodes form the Dirichlet coupling sets $\Gamma_\alpha^n$; coincident coarse- and fine-grid boundary nodes are resolved by the mass-based rule in Eq.~\eqref{eq:overlap_donor}.}
    \label{fig:boundary_node_construction}
\end{figure}

Figure~\ref{fig:boundary_grid_selection} illustrates this particle-based boundary-node selection. Boundary particles scatter through their shape-function supports, and any grid node receiving sufficient contribution from the boundary-particle layer is marked as part of $\Gamma_\alpha^n$.

An additional ambiguity can occur inside $\Omega_{ovlp}$ when both domains identify the same physical location as a boundary grid node. If both sides impose Dirichlet boundary conditions using the other side's velocity, the local velocities can become mutually constrained and stop changing during the Schwarz iteration. We therefore use a mass-based arbitration rule. For each ambiguous overlap node $i$, define
\begin{equation}
    d(i) = \arg\max_{\alpha \in \{B,S\}} m_{\alpha,i}^n,
    \qquad
    r(i) = \{B,S\} \setminus \{d(i)\},
    \label{eq:overlap_donor}
\end{equation}
with ties broken by choosing the background domain $\Omega_B$ as the donor. The domain $d(i)$ with larger local grid mass provides the velocity value, while the other domain $r(i)$ receives that value as its Dirichlet boundary condition. We denote the resulting receiver boundary set after this arbitration by $\widehat{\Gamma}_\alpha^n \subseteq \Gamma_\alpha^n$.

Figure~\ref{fig:boundary_node_ambiguity} illustrates the source of this ambiguity: a coarse-grid boundary node and a fine-grid boundary node may coincide at the same physical location. Once such a coincidence is detected, the mass-based rule in Eq.~\eqref{eq:overlap_donor} assigns only one side as the receiver, avoiding a mutually frozen boundary node during the Schwarz iteration.

It is worth noting that the present interface-node identification is based on the distribution of Lagrangian particles. This provides a flexible way to update the active coupling region and avoids the construction of complex geometric interface data structures for the projection operators. Nevertheless, this particle-based identification may need to be adapted when the material undergoes substantial topological changes, such as fragmentation, separation, merging contact, or highly convoluted self-contact. Examples include large elastoplastic deformation in manufacturing processes and shear localization or failure in geomaterials. In such cases, although the proposed OS-MPM framework is particularly attractive because of its modular refinement capability, the interface-identification strategy should be updated adaptively according to the evolving material configuration. Alternatively, when the critical deformation region can be anticipated from experimental observations or prior simulations, the refined subdomain may be prescribed around the targeted region from the outset.

\subsubsection{Mass-Weighted Spatial Projection ($\Pi$)}
\begin{figure}[htbp]
    \centering
    \includegraphics[width=1.05\textwidth]{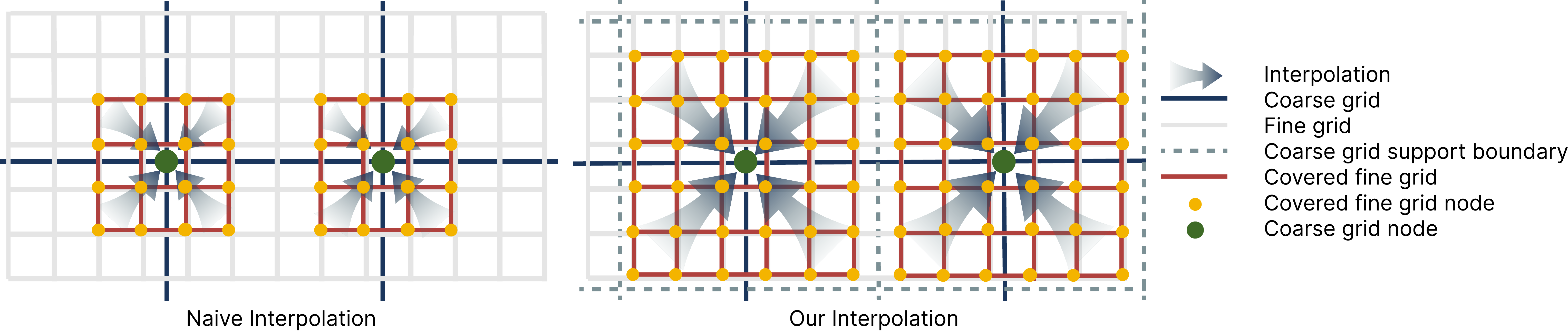}
    \caption{Comparison between interpolation-based fine-to-coarse BC computation (left) and our projection-based version (right).}
    \label{fig:projection_comparison}
\end{figure}

Transferring velocity from the interior of the coarse grid ($I \in \Omega_B$) to a fine grid boundary node ($j \in \widehat{\Gamma}_S^n$) can be directly performed via standard shape function interpolation, denoted $\mathcal{I}_{B \to S}$, because the fine boundary node only requires a pointwise sample of the coarse field:
\begin{equation}
    \mathbf{v}_{S,j} = \mathcal{I}_{B \to S}(\mathbf{v}_B)
    \equiv \sum_{I} \mathbf{v}_{B,I}\, N_{B,I}(\mathbf{x}_{S,j}).
    \label{eq:interp_BtoS}
\end{equation}
The inverse transfer is fundamentally different. Mapping the velocity from the interior of the fine grid ($j \in \Omega_S$) to a coarse grid boundary node ($I \in \widehat{\Gamma}_B^n$) is more of a restriction operation rather than a pointwise sampling. When $\Delta x_B \gg \Delta x_S$, directly interpolating the fine-grid velocity at the coarse nodal position $\mathbf{x}_{B,I}$ would only use a small subset of fine nodes whose fine-grid shape functions are nonzero at that single point. Such a pointwise sampling procedure ignores velocity and momentum contributions carried by other active fine nodes elsewhere inside the support of the coarse basis $N_{B,I}$, and may therefore under-represent the effective coarse-scale state or introduce momentum loss or instabilities. To avoid this issue, we define the projection operator $\Pi_{S \to B}: \mathcal{V}_S \to \mathcal{V}_B$ analogous to the particle-to-grid transfer in MPM as:
\begin{equation}
    \mathbf{v}_{B,I} = \Pi_{S \to B}(\mathbf{v}_S) \equiv \frac{\sum_{j \in \mathcal{N}_S} m_{S,j} \mathbf{v}_{S,j} N_{B,I}(\mathbf{x}_{S,j})}{\sum_{j \in \mathcal{N}_S} m_{S,j} N_{B,I}(\mathbf{x}_{S,j}) + \epsilon_{tol}},
    \label{eq:spatial_projection}
\end{equation}
where $\mathcal{N}_S$ is the set of active fine grid nodes within the support of the coarse node $I$. The numerator and denominator therefore aggregate all active fine-node contributions in the relevant coarse support and overlap region through the weights $m_{S,j} N_{B,I}(\mathbf{x}_{S,j})$, producing a support-wide mass-weighted coarse value rather than a local sample. Fig.~\ref{fig:projection_comparison} illustrates the distinction between a naive pointwise interpolation and the proposed projection. The regularization $\epsilon_{tol}$ ensures numerical stability for nodes with insufficient particle support (vacuum handling). The projection in Eq.~\eqref{eq:spatial_projection} is evaluated only on receiver nodes in $\widehat{\Gamma}_B^n$; outside this set, no Schwarz Dirichlet condition is applied on the coarse grid.

\subsubsection{Temporal Interpolation ($\mathcal{T}$)}
\begin{figure}[htbp]
    \centering
    \includegraphics[width=0.7\textwidth]{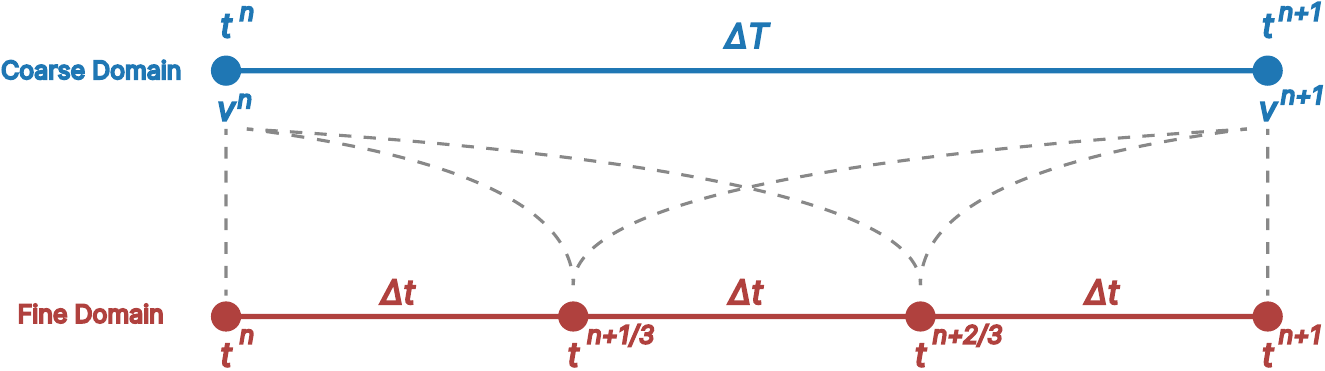}
    \caption{Temporal interpolation used for fine-domain sub-cycling. The coarse-domain interface velocities at $t_n$ and $t_{n+1}$ are linearly interpolated to provide boundary data at each fine sub-step.}
    \label{fig:time_interp}
\end{figure}

The use of different time-step sizes is motivated by the scale separation between the two subdomains \cite{mota2022schwarz,mota2025fundamentally}. Since the fine domain $\Omega_s$ resolves shorter spatial length scales and typically contains the localized deformation, contact, or high-gradient response. It may require a smaller sub-cycles time step for temporal accuracy and, in explicit or weakly damped settings, for stability. In contrast, the coarse domain $\Omega_B$ evolves more smoothly and can therefore be advanced with a larger time step $\Delta T$, avoiding the cost of imposing the fine-scale time step globally. Specifically, for the sub-cycled domain $\Omega_S$, boundary conditions are required at each substep, while the velocity at these intermediate times on $\Omega_B$ is missing. We thus employ a linear temporal interpolation operator $\mathcal{T}_m$ to compute these velocities:
\begin{equation}
    \mathbf{v}_{\Gamma, S}^{n+m/M} = \mathcal{T}_m(\mathbf{v}_{\Gamma, S}^n, \mathbf{v}_{\Gamma, S}^{n+1}) = (1 - \alpha_m)\mathbf{v}_{\Gamma, S}^n + \alpha_m \mathbf{v}_{\Gamma, S}^{n+1}, \quad \text{with } \alpha_m = \frac{m}{M}.
    \label{eq:temporal_interp}
\end{equation}
Figure~\ref{fig:time_interp} illustrates how the coarse-domain interface velocities at $t_n$ and $t_{n+1}$ are linearly interpolated to provide boundary data for the $M$ fine sub-steps.

 The temporal interpolation in Eq.(\ref{eq:temporal_interp}) provides the missing intermediate interface data required by the sub-cycled fine solve. This construction is most accurate when the coarse-interface velocity varies smoothly over $[t_n
,t_{n+1}]$. If the coarse interface undergoes rapid acceleration or unresolved high-frequency motion, an excessively large sub-cycling ratio $M$ may introduce temporal interpolation errors, phase lag, or artificial reflections at the interface. In practice, $M$ should therefore be chosen such that the coarse-domain interface motion remains sufficiently resolved, while the Schwarz residuals provide a direct check of the consistency of the coupled evolution.

\subsection{The Alternating Schwarz Algorithm}
\begin{figure}[htbp]
    \centering
    \includegraphics[width=\textwidth]{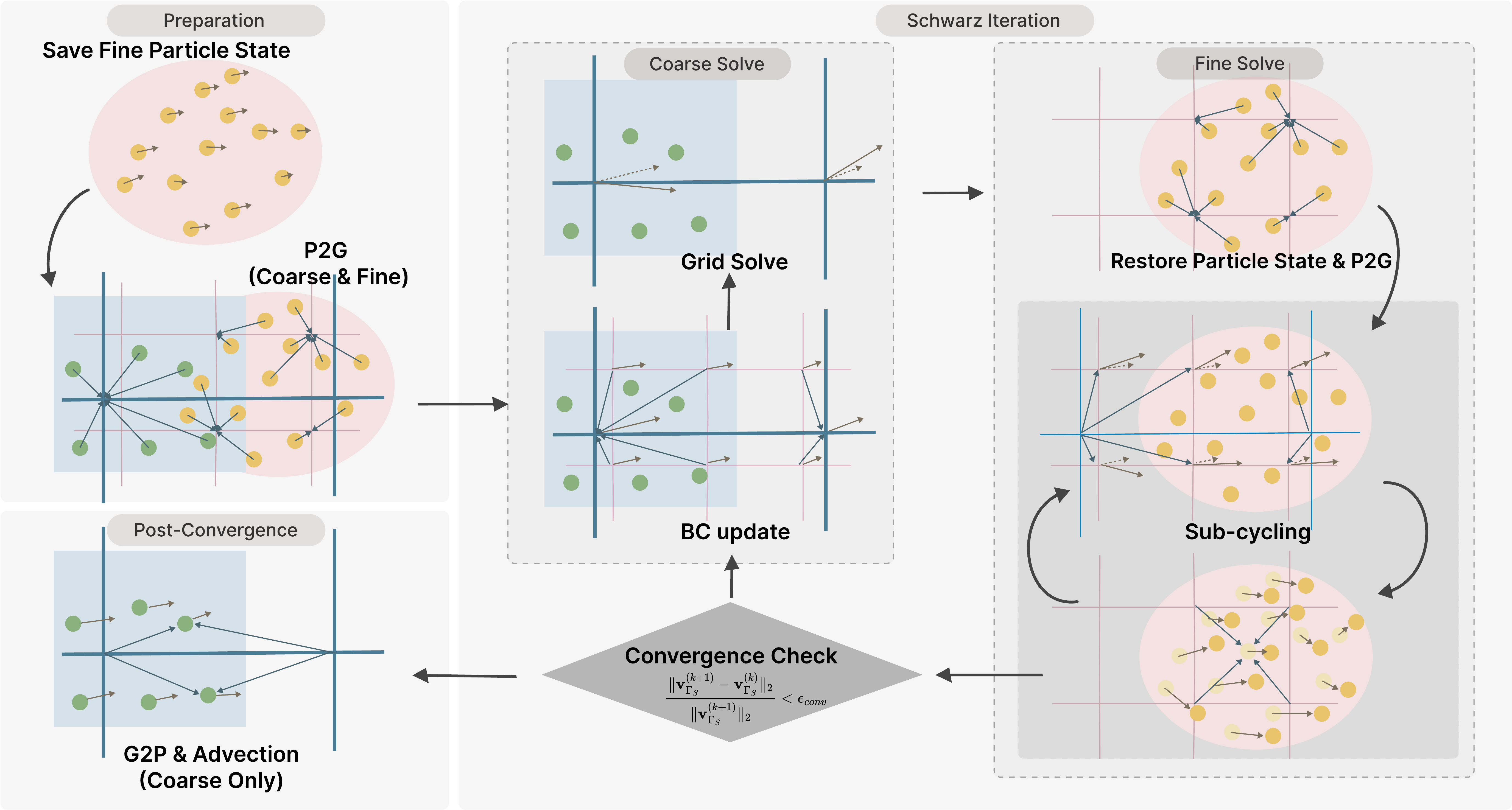}
    \caption{Pipeline overview of our domain-decomposed MPM.}
    \label{fig:pipeline_overview}
\end{figure}

The coupled problem is solved using a multiplicative Schwarz iteration. Let $k$ denote the iteration index. The algorithm seeks the fixed point of the coupled system by alternating between solving $\Omega_B$ and $\Omega_S$ until the interface residuals vanish (see Fig.~\ref{fig:pipeline_overview} and Alg.~\ref{alg:schwarz}).
\subsubsection{Step 0: Preparations before Schwarz Alternating}
Before entering the Schwarz iteration loop, several initialization steps are required. First, the particle state at time $t_n$ is stored for later restoration during each Schwarz iteration: 
\begin{equation}
    \mathcal{S}_S^n \leftarrow \{ \mathbf{x}_{p_S}^n, \mathbf{v}_{p_S}^n, \mathbf{F}_{p_S}^n, \mathbf{B}_{p_S}^n \}_{p_S \in \mathcal{P}_S}.
    \label{eq:store_states}
\end{equation}
Second, an initial grid state is established by performing particle-to-grid (P2G) transfers on both domains, respectively. The boundary grid sets $\Gamma_B^n$ and $\Gamma_S^n$ are then constructed from the boundary-particle P2G contributions using Eq.~\eqref{eq:boundary_grid_set}, followed by the overlap arbitration in Eq.~\eqref{eq:overlap_donor}. The resulting receiver sets $\widehat{\Gamma}_B^n$ and $\widehat{\Gamma}_S^n$ are kept fixed throughout the Schwarz iterations of the current global frame.
Third, the initial guess for the interface boundary condition is set to the particle velocity from the previous time step: $\mathbf{v}_{\Gamma,S}^{(0)} = \mathbf{v}_{\Gamma,S}^n$.

\subsubsection{Step 1: Coarse Problem Solution}
In iteration $k+1$, we solve for $\Omega_B$ over the full step $\Delta T$ using Backward Euler. This is formulated as a constrained minimization:
\begin{equation}
    \mathbf{x}_B^{(k+1)} = \arg \min_{\mathbf{x}} E_B(\mathbf{x})
    \quad \text{subject to } \mathbf{v}_{\widehat{\Gamma}_B^n} = \Pi_{S \to B}(\mathbf{v}_S^{(k)}).
    \label{eq:implicit_solve_coarse}
\end{equation}
The result is a tentative coarse velocity field $\mathbf{v}_B^{(k+1)} = (\mathbf{x}_B^{(k+1)} - \mathbf{x}_B^{n}) / \Delta T$ at $t_{n+1}$.

\subsubsection{Step 2: Fine Problem Sub-cycling}
Solving for $\Omega_S$ involves evolving the fine domain through $M$ sub-steps. A critical implementation detail here is the particle configuration must be reset to the state at $t_n$ at the beginning of each Schwarz iteration to ensure consistent initial conditions.

Before entering the sub-cycling loop, the coarse-domain velocities at $t_n$ and $t_{n+1}$ are mapped onto the selected fine boundary grid $\widehat{\Gamma}_S^n$ via spatial interpolation (Eq.~\ref{eq:interp_BtoS}):
\begin{equation}
    \mathbf{v}_{\Gamma,S}^{n}   = \mathcal{I}_{B \to S}(\mathbf{v}_B^{n}), \quad
    \mathbf{v}_{\Gamma,S}^{n+1} = \mathcal{I}_{B \to S}(\mathbf{v}_B^{(k+1)}). 
    \label{eq:spatial_interp_n}
\end{equation}
These two endpoint velocities are then used inside the loop by the temporal interpolation operator $\mathcal{T}_m$ (Eq.~\ref{eq:temporal_interp} and Fig.~\ref{fig:time_interp}) to supply the boundary condition at each intermediate sub-step.

Before the sub-cycling loop, the fine grid is initialized via particle-to-grid transfer of the reset particle state. Then, for each sub-step $m=1 \dots M$, the procedure involves:
\begin{equation}
    \mathbf{x}_S^{n+m/M} = \arg \min_{\mathbf{x}} E_S(\mathbf{x})
    \quad \text{subject to } \mathbf{v}_{\widehat{\Gamma}_S^n}^{n+m/M} =
    \mathcal{T}_m(\mathbf{v}_{\Gamma,S}^{n}, \mathbf{v}_{\Gamma,S}^{n+1}).
    \label{eq:implicit_solve_fine}
\end{equation}
\begin{enumerate}
    \item \textbf{BC Interpolation:} Compute $\mathbf{v}_{\Gamma,S}^{n+m/M} = \mathcal{T}_m(\mathbf{v}_{\Gamma,S}^n, \mathbf{v}_{\Gamma,S}^{n+1})$ and apply as Dirichlet conditions on the receiver nodes $\widehat{\Gamma}_S^n$.
    \item \textbf{Grid Update (Implicit Solve):} Solve for grid velocities $\mathbf{v}_S^{n+m/M}$ with the constrained update in Eq.~\ref{eq:implicit_solve_fine}.
    \item \textbf{Grid-to-Particle (G2P) Transfer and Particle Advection:} Perform the APIC grid-to-particle transfer to update the particle velocity, affine matrix, and deformation gradient as:
    \begin{align}
        \mathbf{v}_{p_S} &\leftarrow \sum_i N_i(\mathbf{x}_{p_S})\, \mathbf{v}_i, \label{eq:g2p_v_substep}\\
        \mathbf{B}_{p_S} &\leftarrow \sum_i N_i(\mathbf{x}_{p_S})\, \mathbf{v}_i (\mathbf{x}_i - \mathbf{x}_{p_S})^T, \label{eq:g2p_B_substep}\\
        \mathbf{F}_{p_S} &\leftarrow \left(\mathbf{I} + \Delta t \sum_i \mathbf{v}_i
    \left(\nabla^{\mathbf{x}} N_i(\mathbf{x}_{p_S})\right)^T \right) \mathbf{F}_{p_S}, \label{eq:F_substep}
    \end{align}
    followed by particle advection:
    \begin{align}
        \mathbf{x}_{p_S} &\leftarrow \mathbf{x}_{p_S} + \Delta t \sum_i \mathbf{v}_i N_i(\mathbf{x}_{p_S}), \label{eq:advect_substep}
    \end{align}
    consistent with Eqs.~\ref{eq:F_update}-\ref{eq:advection}.
    \item \textbf{Particle-To-Grid (P2G) Transfer:} Particle mass and momentum are transferred to the grid nodes using the APIC scheme (Section~\ref{subsubsec:mpm_particles}), preparing the grid for the next sub-step:
    \begin{equation}
    \begin{aligned}
    m_i & = \sum_{p_S} m_{p_S} N_i(\mathbf{x}_{p_S}), \\
        (m\mathbf{v})_i & = \sum_{p_S} m_{p_S} N_i(\mathbf{x}_{p_S}) \left(\mathbf{v}_{p_S} + \mathbf{B}_{p_S} (\mathbf{D}_{p_S})^{-1}(\mathbf{x}_i - \mathbf{x}_{p_S})\right).
    \end{aligned}
        \label{eq:p2g_substep}
    \end{equation}
\end{enumerate}

\begin{algorithm}[H]
\caption{Domain-Decomposed MPM via Schwarz Alternating Method}
\label{alg:schwarz}
\begin{algorithmic}[1]
\State \textbf{Input:} State at $t_n$, Tolerance $\epsilon_{conv}$.

\vspace{0.1cm}
\State \textit{// 0. Preparations before Schwarz Alternating}
\State Store fine particle states (Eq.~\ref{eq:store_states}).
\State Particle-to-Grid Transfer on $\Omega_B$ and $\Omega_S$ (Eq.~\ref{eq:p2g_transfer}).
\State Identify boundary grid sets $\Gamma_B^n$ and $\Gamma_S^n$ from boundary particles and apply overlap arbitration (Eqs.~\ref{eq:boundary_grid_set}--\ref{eq:overlap_donor}).

\For{iteration $k = 0 \to k_{max}$}
    \State \textit{// 1. Solve Coarse Domain}
    \State Compute interface BCs on $\widehat{\Gamma}_B^n$ (Eq.~\ref{eq:spatial_projection}).
    \State Implicit Solve on $\Omega_B$ for $\Delta T$ (Eq.~\ref{eq:implicit_solve_coarse}).

    \vspace{0.1cm}
    \State \textit{// 2. Solve Fine Domain (Sub-cycling)}
    \State \label{algline:reset} Restore fine particle states.
    \State  Particle-to-Grid Transfer on $\mathcal{P}_S$ (Eq.~\ref{eq:p2g_substep}).
    \State Compute interface BCs on $\widehat{\Gamma}_S^n$ (Eq.~\ref{eq:spatial_interp_n}).
    \For{sub-step $m = 1 \to M$}
        \State Temporally Interpolate BCs (Eq.~\ref{eq:temporal_interp}).
        \State Implicit Solve on $\Omega_S$ for $\Delta t$ (Eq.~\ref{eq:implicit_solve_fine}).
        \State Grid-to-Particle Transfer on $\Omega_S$ (Eqs.~\ref{eq:g2p_v_substep}--\ref{eq:F_substep}).
        \State Advection on $\Omega_S$ (Eq.\ref{eq:advect_substep}).
        \State Particle-to-Grid Transfer on $\Omega_S$ (Eq.~\ref{eq:p2g_substep}).
    \EndFor

    \vspace{0.1cm}
    \State \textit{// 3. Check Convergence}
\If{converged (Eqs.~\ref{eq:convergence_residuals}--\ref{eq:convergence_criterion})} 
\State \textbf{Break}; 
\EndIf
\EndFor

\vspace{0.1cm}
\State \textit{// 4. Post-Convergence: Update Coarse Particles $\mathcal{P}_B$}
\State Grid-to-Particle Transfer on $\Omega_B$ (Eqs.~\ref{eq:F_update}--\ref{eq:g2p_B}).
\State Advection on $\Omega_B$ (Eqs.~\ref{eq:advection}).

\vspace{0.1cm}
\State \textbf{Output:} State at $t_{n+1}$. // $\mathcal{P}_S$ already advanced; $\mathcal{P}_B$ just updated.
\end{algorithmic}
\end{algorithm}
\paragraph{Remark on Consistency of Interface Transfer}
\textbf{APIC} is used to guarantee exact angular momentum conservation during interface transfers; otherwise, the disparate spatial resolutions of $\Omega_B$ and $\Omega_S$ would dissipate angular momentum at mismatched rates, introducing unphysical parasitic torques.


\subsection{Convergence Criterion}
The iteration is terminated when the interface velocity fields on \emph{both} grids stabilize between consecutive Schwarz iterates. Letting $\mathbf{v}_{\Gamma_B}^{(k)} = \Pi_{S \to B}(\mathbf{v}_S^{(k)})$ denote the projected velocity applied as the Dirichlet BC on $\widehat{\Gamma}_B^n$, and $\mathbf{v}_{\Gamma_S}^{(k)} = \mathcal{I}_{B \to S}(\mathbf{v}_B^{(k)})$ denote the interpolated velocity applied on $\widehat{\Gamma}_S^n$, we define the two interface residuals:
\begin{equation}
    r_B^{(k)} = \frac{\| \mathbf{v}_{\Gamma_B}^{(k+1)} - \mathbf{v}_{\Gamma_B}^{(k)} \|_2}{\| \mathbf{v}_{\Gamma_B}^{(k+1)} \|_2}, \qquad
    r_S^{(k)} = \frac{\| \mathbf{v}_{\Gamma_S}^{(k+1)} - \mathbf{v}_{\Gamma_S}^{(k)} \|_2}{\| \mathbf{v}_{\Gamma_S}^{(k+1)} \|_2}.
    \label{eq:convergence_residuals}
\end{equation}
Convergence is declared when both residuals fall below the tolerance $\epsilon_{conv}$:
\begin{equation}
    r_B^{(k)} \le \epsilon_{conv} \quad \text{and} \quad r_S^{(k)} \le \epsilon_{conv}.
    \label{eq:convergence_criterion}
\end{equation}

\subsection{Post-Convergence State Update}
\label{subsec:post_convergence}
Upon convergence at iteration $k^*$, the two particle sets require different treatment to advance to state $t_{n+1}$. \textbf{$\Omega_S$ particles ($\mathcal{P}_S$)} are already at state $t_{n+1}$. The final sub-step of iteration $k^*$ completed the full G2P and advection sequence (Eqs.~\ref{eq:g2p_v_substep}--\ref{eq:advect_substep}), and the particles are \emph{not} reset after the convergence check and no further action is needed. \textbf{$\Omega_B$ particles ($\mathcal{P}_B$)} remained at state $t_n$ throughout the Schwarz iterations, because the coarse domain solve only produces a grid velocity $\mathbf{v}_B^{n+1} \equiv \mathbf{v}_B^{(k^*+1)}$ without updating the particles. Therefore, a single post-convergence APIC G2P transfer and particle advection are applied to advance $\mathcal{P}_B$, following Eqs.~\ref{eq:F_update}--\ref{eq:advection}.

%% file: sections/results.tex
\section{Numerical Examples}
\label{sec:numerical_examples}

In this section, we present three numerical benchmarks and a brief three-dimensional showcase for the proposed OS-MPM. All simulations are implemented in Taichi\footnote{https://www.taichi-lang.org/}. The three benchmarks evaluate the method's accuracy, convergence, and computational efficiency. We begin with two unit tests: a large-deformation cantilever beam and a Hertzian contact problem, which validate the method's ability to handle severe geometric nonlinearities and curved contact boundaries. We then present the elastic inclusion problem, which provides both a convergence under refinement validation against an analytical solution and a computational efficiency comparison. Finally, we include a compact three-dimensional showcase to demonstrate the natural extension of the proposed framework to 3D settings, where its modular coarse–fine coupling and local space–time refinement capabilities are particularly relevant for engineering applications involving localized deformation.

Unless otherwise specified, we employ a Neo-Hookean hyperelastic material model for all simulations. The Schwarz iteration tolerance is set to $\epsilon_{conv} = 10^{-5}$. All computational performance benchmarks are evaluated on a workstation equipped with a Ryzen 9950X3D CPU and 64 GB of RAM.
\subsection{Cantilever Beam}
\label{subsec:exp_cantilever}
In the first experiment, we adopt the gravity-driven cantilever benchmark of Romero et al.~\cite{bickley1934,shield1992} to verify that our formulation for OS-MPM does not introduce artificial stiffness or locking under extreme geometric nonlinearity. A slender clamped-free beam deforms quasi-statically in a plane under its own weight, so the equilibrium is controlled by the competition between gravitational loading and bending rigidity. For a rectangular cross-section, this balance is characterized by the gravito-bending length
\begin{equation}
    L_{\mathrm{gb}}^* = \left(\frac{Dw}{\rho A^* g}\right)^{1/3},
    \qquad
    \Gamma_{\mathrm{gb}}^* = \left(\frac{L}{L_{\mathrm{gb}}^*}\right)^3
    = \frac{\rho A^* g L^3}{Dw},
\end{equation}
where $A^* = wh$ and $D = Eh^3/[12(1-\nu^2)]$ is the bending stiffness of the beam. Consequently, any set of dimensional parameters sharing the same $\Gamma_{\mathrm{gb}}^*$ must collapse onto the same equilibrium response. In our simulations we fix $E = 10^2\,\mathrm{kPa}$, $\nu = 0.4$, and $\rho = 1000\,\mathrm{kg/m^3}$, and vary the gravito-bending parameter $\Gamma_{\mathrm{gb}}^*$ by adjusting the gravitational acceleration $g$. The spatial grid spacing is $h = 5\times 10^{-3}\,\mathrm{m}$, the time step is $\Delta t = 2\times 10^{-3}\,\mathrm{s}$, and each grid cell is initialized with $9$ particles.

\figCantileverSetup

Following \citet{bickley1934,shield1992}, the planar equilibrium is governed by the elastica boundary value problem
\begin{equation}
\begin{aligned}
    \frac{d^2\theta}{d\bar{s}^2} + \Gamma_{\mathrm{gb}}^*(1-\bar{s})\cos\theta &= 0,
    \qquad \theta(0) = 0,\quad \frac{d\theta}{d\bar{s}}(1) = 0,\\
    \frac{d\bar{x}}{d\bar{s}} &= \cos\theta,\qquad \bar{x}(0) = 0,\\
    \frac{d\bar{y}}{d\bar{s}} &= \sin\theta,\qquad \bar{y}(0) = 0,
\end{aligned}
\end{equation}
where $\bar{s}=s/L$ and $(\bar{x}, \bar{y}) = (x/L, y/L)$. This dimensionless formulation implies that the aspect ratio $H/W$ of the equilibrium shape is a unique monotone function of $\Gamma_{\mathrm{gb}}^*$, which makes the cantilever test a compact master-curve benchmark for large-deformation bending.

\begin{figure}[htbp]
    \centering
    \includegraphics[width=0.6\textwidth]{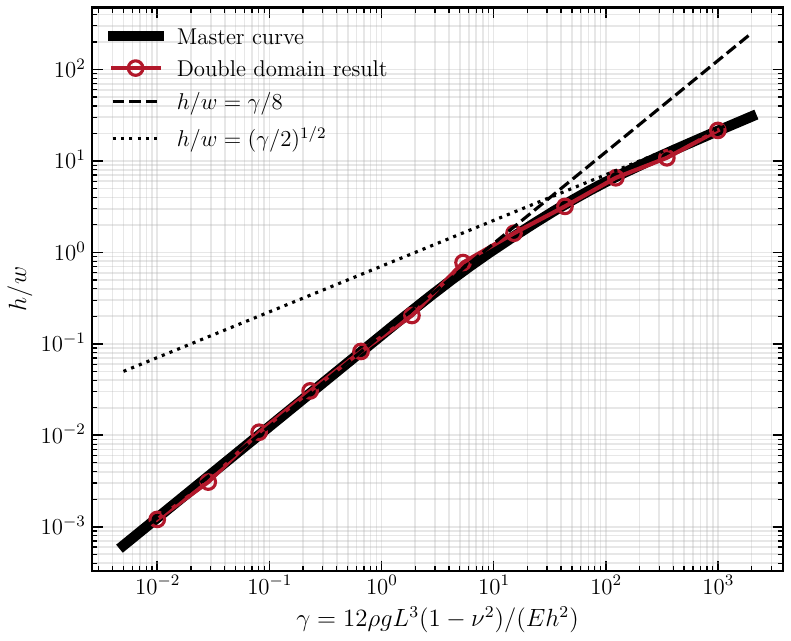}
    \caption{Master-curve validation for the gravity-driven cantilever beam experiment. Our OS-MPM results agree with the analytical planar-elastica relation between the deformation aspect ratio $H/W$ and the gravito-bending parameter $\Gamma_{\mathrm{gb}}^*$.}
    \label{fig:cantilever_trajectory}
\end{figure}

In our dual-domain setting, the beam is split longitudinally into two overlapping subdomains, so this test directly probes whether the Schwarz coupling preserves the correct effective bending response. Fig.~\ref{fig:cantilever_trajectory} shows the $\Gamma_{\mathrm{gb}}^*$--$H/W$ master curve computed with OS-MPM and compared against the analytical planar-elastica solution. The numerical results closely match the analytical curve across the tested deformation range. 

\begin{figure}[htbp]
    \centering
    \includegraphics[width=0.68\textwidth]{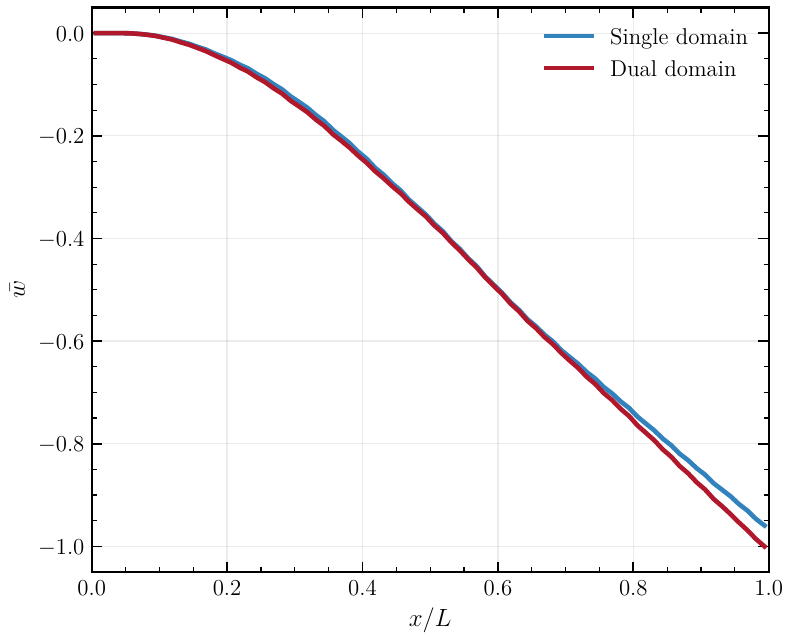}
    \caption{Normalized displacement comparison for the cantilever beam at $\Gamma_{\mathrm{gb}}^* = 2.31\times 10^{-1}$. The single-domain and dual-domain results nearly overlap.}
    \label{fig:cantilever_displacement_comparison}
\end{figure}

This master-curve comparison is complemented by a representative normalized displacement profile at $\Gamma_{\mathrm{gb}}^* = 2.31\times 10^{-1}$. In Fig.~\ref{fig:cantilever_displacement_comparison}, both axes are normalized: the horizontal coordinate is given by $x/L$, and the displacement is shown in normalized form as $\bar{u}$. The single-domain and dual-domain curves nearly overlap throughout the beam, confirming that the Schwarz coupling preserves the global bending response. The remaining small discrepancy can be attributed to the non-monolithic nature of the Schwarz coupling, where the two subdomains exchange kinematic information through interpolation/projection over the overlap region. In this case, the interface compatibility is satisfied only up to the prescribed Schwarz tolerance. In addition, the independent particle–grid transfers in the two subdomains introduce slightly different quadrature and interpolation errors near the overlap. However, these effects remain localized and do not lead to any visible artificial stiffening or distortion of the overall beam deformation.

\subsection{Hertzian Contact Problem}
\label{subsec:exp_hertz}
The Hertzian contact problem serves as a standard benchmark for assessing the resolution of localized contact tractions. Under frictionless normal contact, the pressure field is confined to a narrow contact patch and exhibits strong gradients near the rigid boundary. The problem is therefore well suited for evaluating local refinement strategies, since the relevant length scales are highly localized while the far-field deformation remains comparatively smooth.

\figHertzContact

We consider a two-dimensional compliant semi-circular cylinder of radius $R$ pressed against a rigid horizontal plane. The cylinder is modeled with Young's modulus $E = 2\times 10^2\,\mathrm{kPa}$, Poisson's ratio $\nu = 0.3$, and density $\rho = 1000\,\mathrm{kg/m^3}$. Since the steepest stress variation is confined to the lower contact region, accurate resolution of the contact patch requires substantially finer discretization near the interface than in the bulk. We therefore place the fine subdomain $\Omega_S$ over the anticipated contact zone and use the coarse subdomain $\Omega_B$ to represent the remaining bulk response. The coarse domain uses a fixed grid spacing $h_B = 1.56 \times 10^{-2}\,\mathrm{m}$ and time step $\Delta T = 2\times 10^{-4}\,\mathrm{s}$. The fine-domain grid spacing is refined from $h_S = 1.5\times 10^{-2}\,\mathrm{m}$ to $5\times 10^{-3}\,\mathrm{m}$, with the fine time step ranging from $\Delta t = 2\times 10^{-4}\,\mathrm{s}$ to $5\times 10^{-5}\,\mathrm{s}$; each grid cell is initialized with $16$ particles. This decomposition directly tests whether the overlapping Schwarz coupling can recover the local contact solution without resorting to uniform global refinement.

The problem is governed by the classical Hertz contact theory~\cite{johnson1985contact}. Assuming a normal applied force $F$ per unit length, the contact pressure profile $p(x)$ over the contact interface is distributed semi-elliptically:
\begin{equation}
    p(x) = p_{\text{max}} \sqrt{1 - \left(\frac{x}{b}\right)^2},
\end{equation}
where $p_{\text{max}} = 2F / (\pi b)$ is the maximum contact pressure at the center ($x=0$), and $b$ denotes the contact half-width. For a cylinder with Young's modulus $E$ and Poisson's ratio $\nu$ contacting a rigid half-space, the half-width is given by:
\begin{equation}
    b = 2\sqrt{\frac{F R (1-\nu^2)}{\pi E}}.
\end{equation}
These expressions define both the target pressure profile and the corresponding contact half-width, and thus provide a stringent quantitative benchmark for the dual-domain discretization.

\begin{figure}[htbp]
    \centering
    \includegraphics[width=0.9\textwidth]{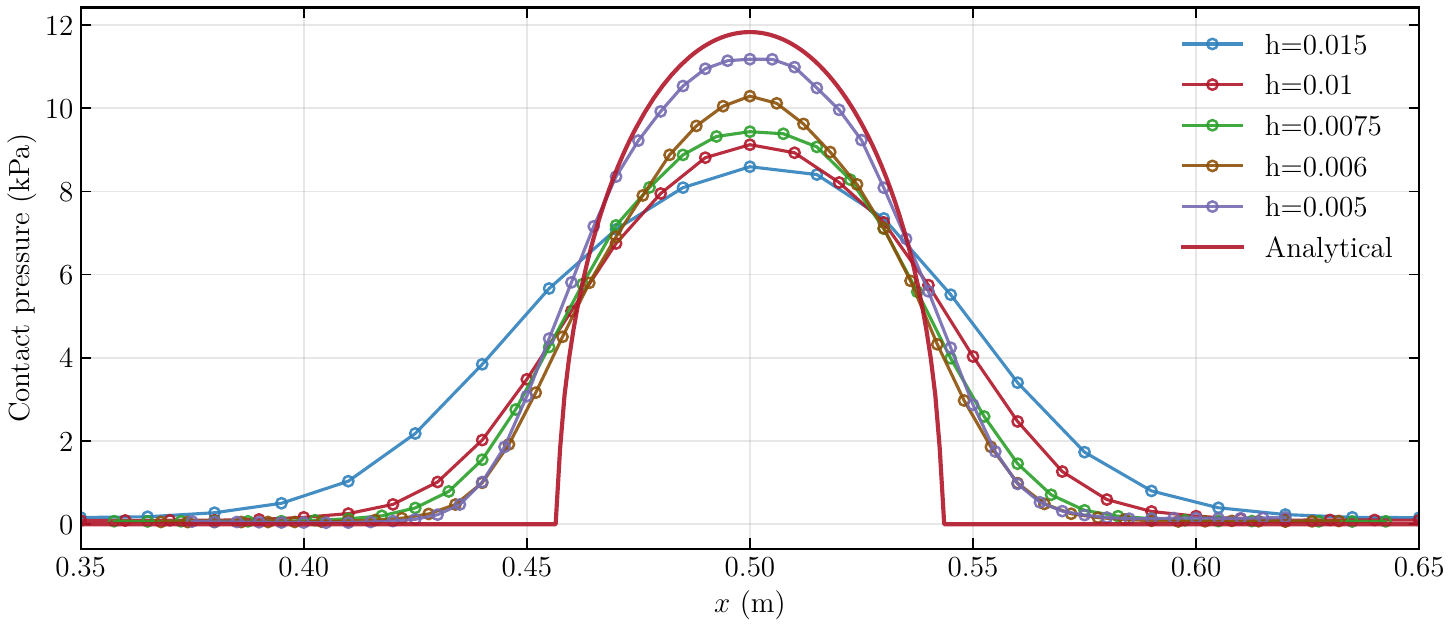}
    \caption{Hertzian contact benchmark. Contact pressure distribution along the rigid interface for a sequence of fine-domain ($\Omega_S$) resolutions, with the coarse domain ($\Omega_B$) held fixed. The numerical profiles converge monotonically toward the analytical Hertz solution as the fine-grid mesh is refined.}
    \label{fig:hertz_pressure}
\end{figure}

Fig.~\ref{fig:hertz_pressure} compares the extracted numerical contact pressure profiles with the analytical Hertz solution across a sequence of meshes. The coarse domain $\Omega_B$ is held at a fixed moderate resolution throughout, while the fine domain $\Omega_S$ is progressively refined. As the fine-grid resolution increases, the numerical pressure profile converges monotonically toward the analytical solution, with the error reducing approximately linearly with mesh element size. OS-MPM consistently reproduces both the pressure peak and the contact half-width, and the residual deviation near the contact edge decreases with refinement, consistent with the particle-quadrature and diffuse-boundary effects inherent to MPM. These results demonstrate that selective refinement of $\Omega_S$ alone is sufficient to drive convergence of the contact response, while the overlapping Schwarz coupling ensures an accurate transfer of kinematics to the coarse far field.

\subsection{Elastic Inclusion Problem}

\label{subsec:exp_inclusion}
The previous two benchmarks assess the ability of the proposed framework to handle large deformation and localized contact. We next consider a material-heterogeneity problem to examine whether the Schwarz coupling can accurately transmit stress fields across a coarse–fine decomposition when sharp gradients arise from internal material contrast rather than external boundary loading. This benchmark also provides an analytical reference solution, allowing both convergence under refinement and computational efficiency to be quantified. Specifically, we consider the classical Eshelby inclusion problem~\cite{eshelby1957,mura2013}: an elastic matrix containing a circular inclusion that carries a prescribed misfit eigenstrain. The misfit is introduced numerically by initializing the deformation gradient of all inclusion particles to
\begin{equation}
    \mathbf{F}_0 = (1 - \delta)\,\mathbf{I},
\end{equation}
where $\delta > 0$ is the compression parameter. For this equi-axial contraction, the Green strain is $\mathbf{E} = \frac{1}{2}(\mathbf{F}_0^T\mathbf{F}_0 - \mathbf{I}) = \frac{1}{2}((1-\delta)^2 - 1)\mathbf{I} = \left(\frac{\delta^2}{2} - \delta\right)\mathbf{I}$. In the linear strain limit ($\delta \ll 1$), this simplifies to $\mathbf{E} \approx -\delta\,\mathbf{I}$, corresponding to an isotropic compressive eigenstrain $\varepsilon^* \approx -\delta$. Throughout the following analysis, we employ this linear-strain approximation and denote the effective strain magnitude as $\delta_{\text{eff}} = \delta$. The system is then allowed to relax to static equilibrium. A key feature of this problem is that the exact solution is piecewise analytic: the stress state inside the inclusion is uniform (hydrostatic), while outside it decays as $(R/r)^2$. The resulting steep stress gradient at the inclusion boundary makes this a stringent benchmark for spatial resolution.

\figInclusionSetup

Under the plane strain assumption, the analytical solution can be derived from classical micromechanics~\cite{mura2013} using the Lamé formulas. Let $G = E / [2(1+\nu)]$ denote the shear modulus. The uniform contact pressure $P$ at the inclusion-matrix interface is determined by the compatibility condition:
\begin{equation}
    P = \frac{\delta_{\text{eff}}}{C_{\text{out}} + C_{\text{in}}},
\end{equation}
where $C_{\text{out}} = 1 / (2G_{\text{out}})$ and $C_{\text{in}} = (1 - 2\nu_{\text{in}}) / (2G_{\text{in}})$ are the compliances of the exterior matrix and the interior inclusion, respectively. The exact radial stress ($\sigma_{rr}$) and transverse hoop stress ($\sigma_{\theta\theta}$) fields in polar coordinates $(r, \theta)$ are given piecewise. Inside the inclusion ($r < R$), the stress state is uniformly compressive:
\begin{equation}
    \sigma_{rr} = -P, \quad \sigma_{\theta\theta} = -P.
\end{equation}
Outside the inclusion ($r \ge R$), the stresses decay quadratically with respect to the radial distance:
\begin{equation}
    \sigma_{rr} = -P \left(\frac{R}{r}\right)^2, \quad \sigma_{\theta\theta} = P \left(\frac{R}{r}\right)^2.
\end{equation}

We employ our domain decomposition framework by assigning the fine-grid domain ($\Omega_S$) to the circular inclusion and its immediate vicinity, where the stress gradient is concentrated. The smoothly decaying far-field response is covered by the coarse-grid domain ($\Omega_B$). The matrix (exterior domain $\Omega_B$) has Young's modulus $E_{\mathrm{out}} = 1.0\times 10^2\,\mathrm{kPa}$ and Poisson's ratio $\nu_{\mathrm{out}} = 0.3$ throughout all cases. The inclusion (fine domain $\Omega_S$) shares $\nu_{\mathrm{in}} = 0.3$, with $E_{\mathrm{in}}$ determined by the stiffness ratio $E_{\mathrm{in}}/E_{\mathrm{out}}$. We test three stiffness ratios $E_{\mathrm{in}}/E_{\mathrm{out}} \in \{1, 2, 5\}$ under a sequence of successively refined meshes. For the single-domain reference, the coarsest grid uses a cell size $h$ with time step $\Delta t = 0.01\,\mathrm{s}$; finer meshes keep the ratio $\Delta t / h$ constant, so the time step decreases proportionally with grid spacing. For the dual-domain simulations, the fine domain $\Omega_S$ is always discretized at the same resolution as the corresponding single-domain case, while the coarse domain $\Omega_B$ uses a cell size twice as large ($2h$). The coarse time step satisfies $\Delta T = 2\Delta t$, matching the sub-cycling ratio $M = 2$ throughout all refinement levels. Each grid cell is initialized with $4$ particles. Simulations are run to static equilibrium; the stopping criterion requires the $\ell^\infty$ norm of the grid velocity to fall below $10^{-6}$ on \emph{both} grids in OS-MPM.

\begin{figure}[htbp]
    \centering
        \includegraphics[width=\textwidth]{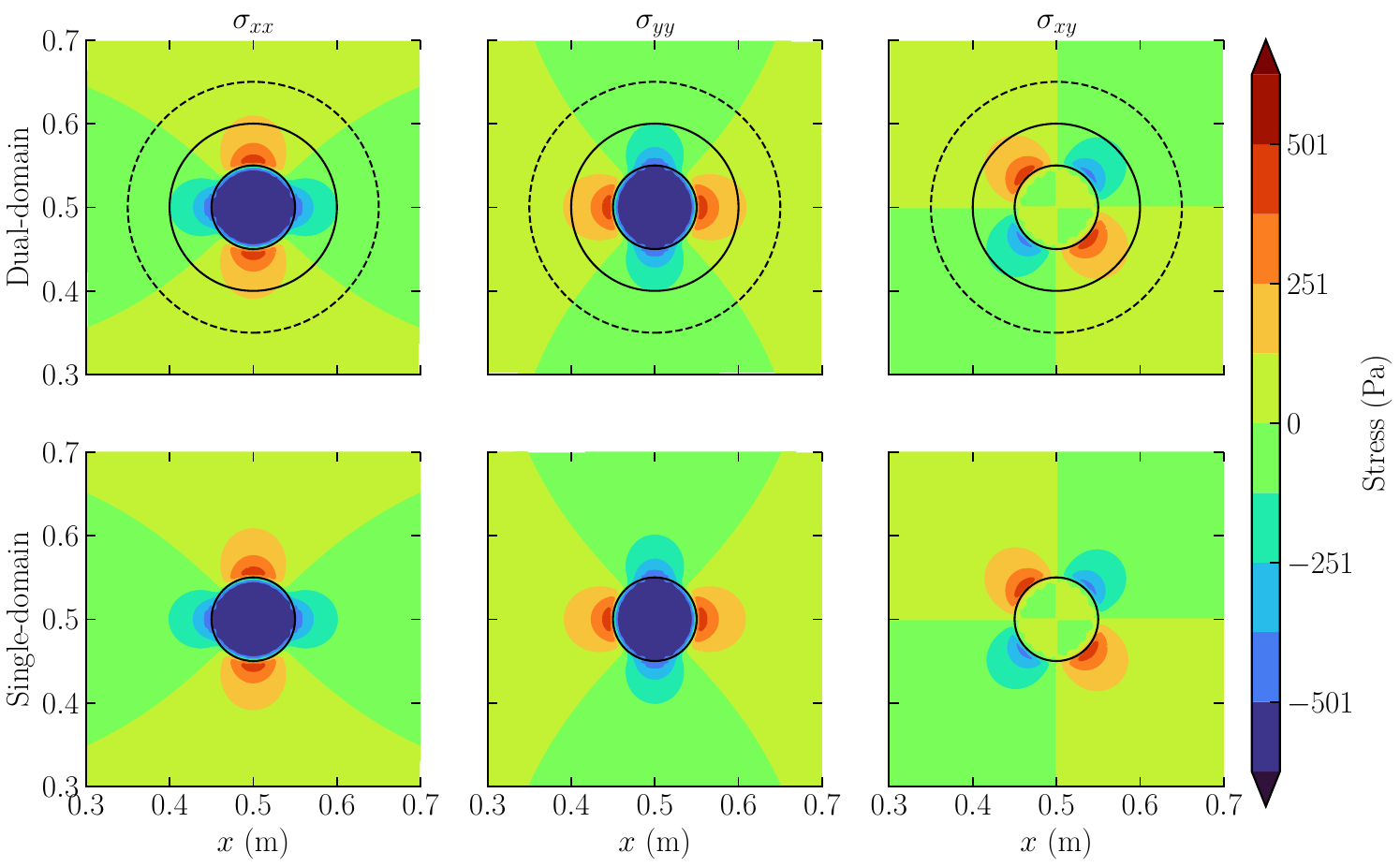}
    \caption{Stress distribution for the elastic inclusion problem ($h = 0.004\,\mathrm{m}$) for stiffness ratio $E_\mathrm{in}/E_\mathrm{out} = 1$. Each panel: top row is dual-domain Schwarz MPM; bottom row is single-domain reference. Columns show $\sigma_{xx}$, $\sigma_{yy}$, and $\sigma_{xy}$, respectively. Solid and dashed black circles indicate the coarse-domain ($\Omega_B$) and fine-domain ($\Omega_S$) boundaries.}
    \label{fig:stress_dist_g100_1}
\end{figure}

\begin{figure}[htbp]
    \centering
        \includegraphics[width=0.95\textwidth]{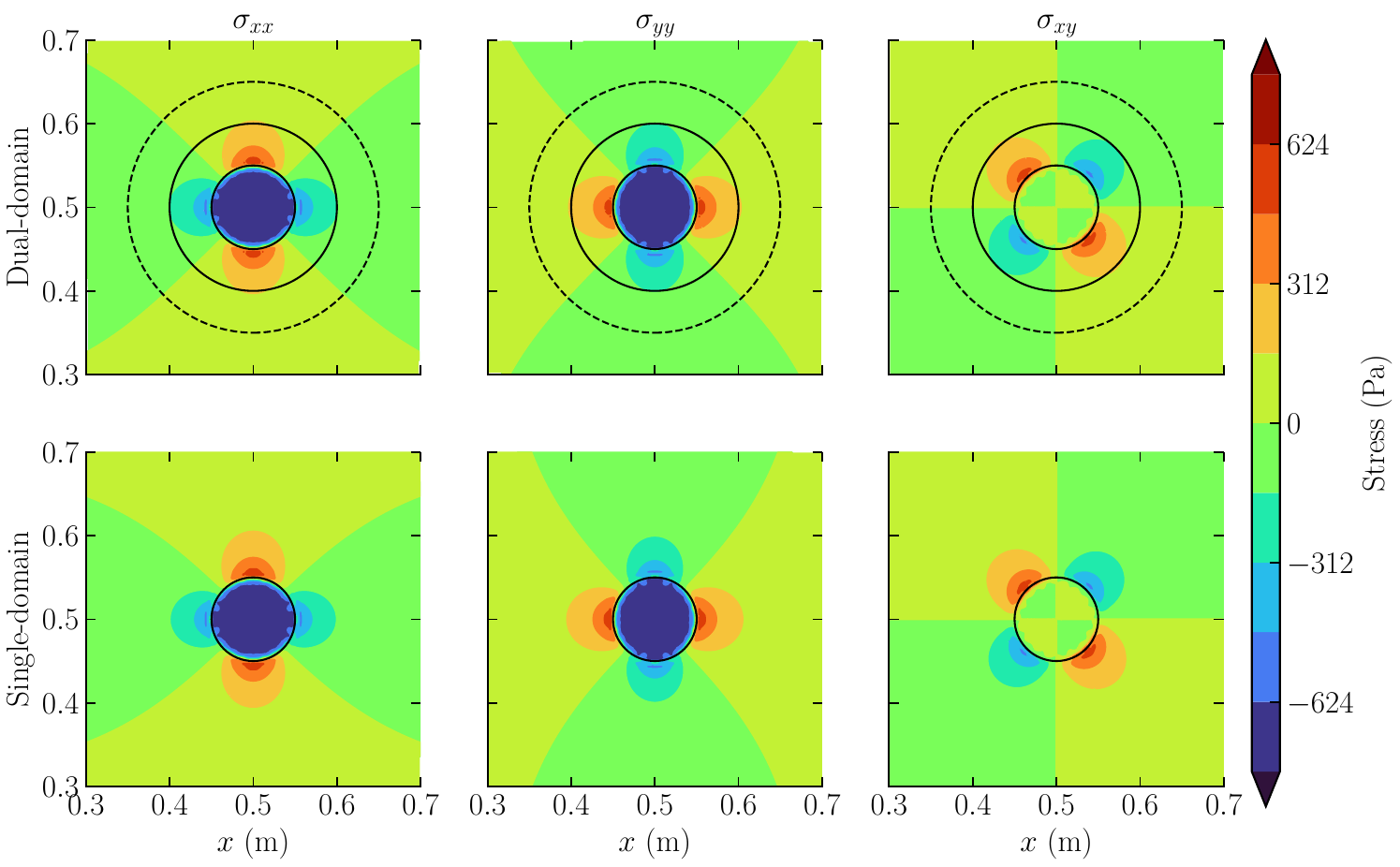}
    \caption{ $E_\mathrm{in}/E_\mathrm{out} = 2$.}
    \label{fig:stress_dist_g100_2}
\end{figure}

\begin{figure}[htbp]
    \centering
        \includegraphics[width=0.95\textwidth]{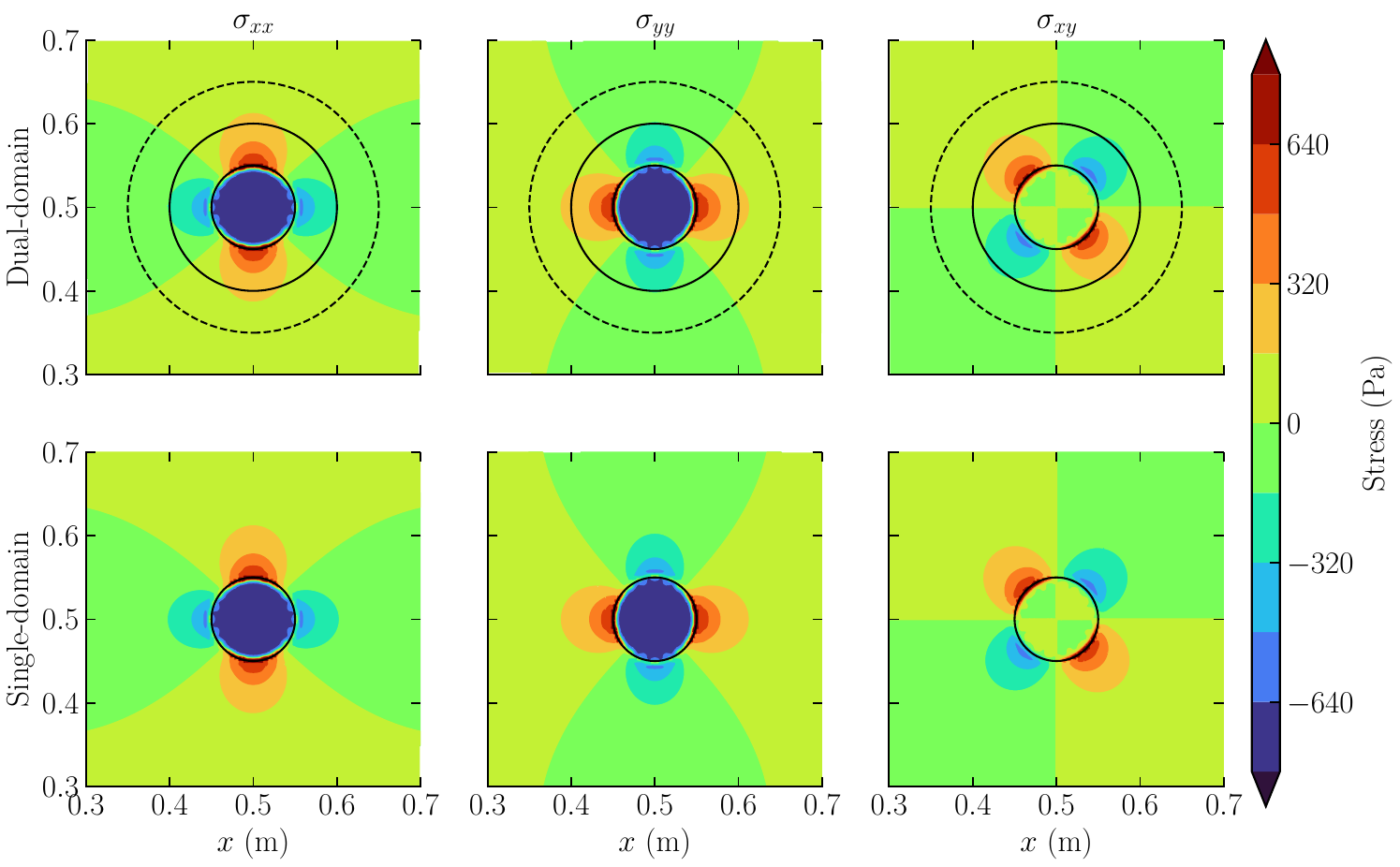}
    \caption{$E_\mathrm{in}/E_\mathrm{out} = 5$.}
    \label{fig:stress_dist_g100_5}
\end{figure}

The \emph{stress distributions} in Figs.~\ref{fig:stress_dist_g100_1}, \ref{fig:stress_dist_g100_2}, and \ref{fig:stress_dist_g100_5} show the full $(\sigma_{xx}, \sigma_{yy}, \sigma_{xy})$ stress fields at the finest common resolution ($h = 4.0\times10^{-3}\,\mathrm{m}$) for all three stiffness ratios. In each panel the dual-domain result occupies the top row and the single-domain reference the bottom row. Across all stiffness cases and all three stress components, the two solutions are visually indistinguishable, confirming that the Schwarz interface introduces no spurious stress artifacts regardless of the material contrast. The characteristic features of the analytical solution—uniform hydrostatic compression inside the inclusion and the $1/r^2$ decay outside—are faithfully reproduced by both solvers.

The localized discrepancy near the inclusion boundary, particularly in the higher stiffness-contrast cases in Figs.~\ref{fig:stress_dist_g100_2} and \ref{fig:stress_dist_g100_5}, is mainly associated with the representation of a sharp material interface on a structured MPM background grid. The inclusion boundary introduces a weak discontinuity: the displacement field remains continuous, while strain and stress may change abruptly across the material interface. Since the circular interface is not grid-conforming, particle quadrature and grid interpolation smear this transition over a narrow region, producing localized stress oscillations or singularity-like artifacts. These errors are expected to become more pronounced as the stiffness contrast increases. Although they remain confined to the interface region in the present benchmark, problems requiring highly smooth or accurate interface stresses may benefit from additional weak-discontinuity treatments, such as enriched shape functions or interface-conforming discretizations.


\begin{figure}[htbp]
    \centering
    \begin{subfigure}[b]{0.41\textwidth}
        \includegraphics[width=\textwidth]{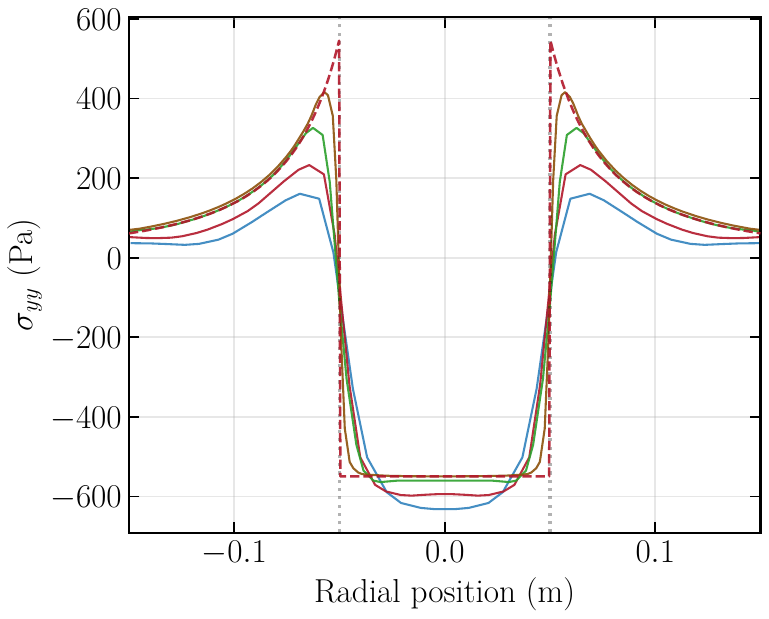}
        \caption{Dual-domain, $E_\mathrm{in}/E_\mathrm{out}=1$}
        \label{fig:summary_yy_1_double}
    \end{subfigure}\hfill
    \begin{minipage}[b]{0.15\textwidth}
        \centering
        \vspace{0.6cm}
        \includegraphics[width=1.0\textwidth]{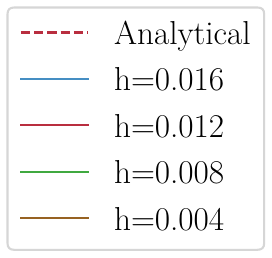}
    \end{minipage}\hfill
    \begin{subfigure}[b]{0.41\textwidth}
        \includegraphics[width=\textwidth]{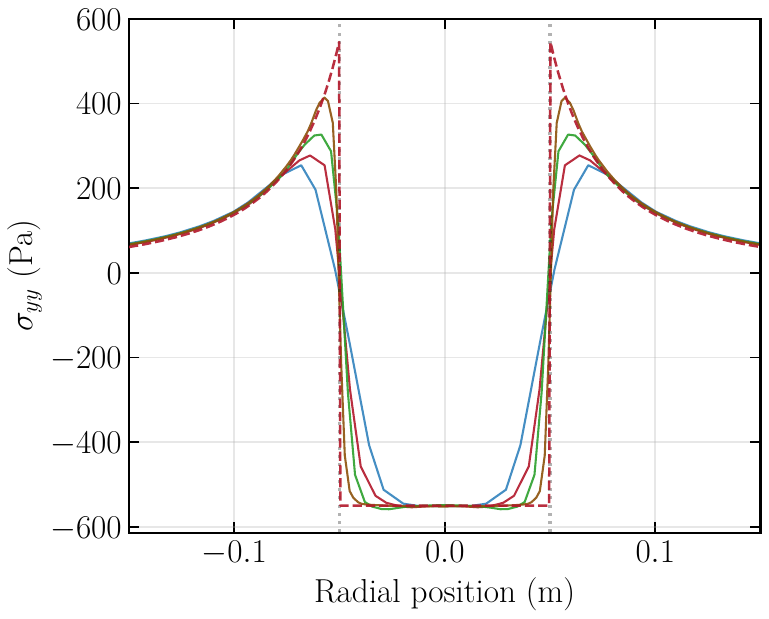}
        \caption{Single-domain, $E_\mathrm{in}/E_\mathrm{out}=1$}
        \label{fig:summary_yy_1_single}
    \end{subfigure}
    \par\vspace{0.12cm}

    \begin{subfigure}[b]{0.41\textwidth}
        \includegraphics[width=\textwidth]{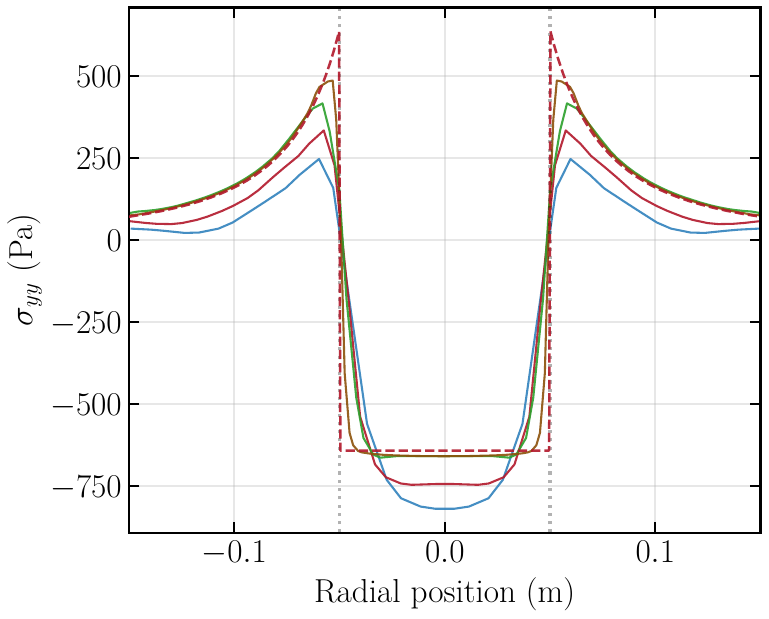}
        \caption{Dual-domain, $E_\mathrm{in}/E_\mathrm{out}=2$}
        \label{fig:summary_yy_2_double}
    \end{subfigure}\hfill
    \begin{subfigure}[b]{0.41\textwidth}
        \includegraphics[width=\textwidth]{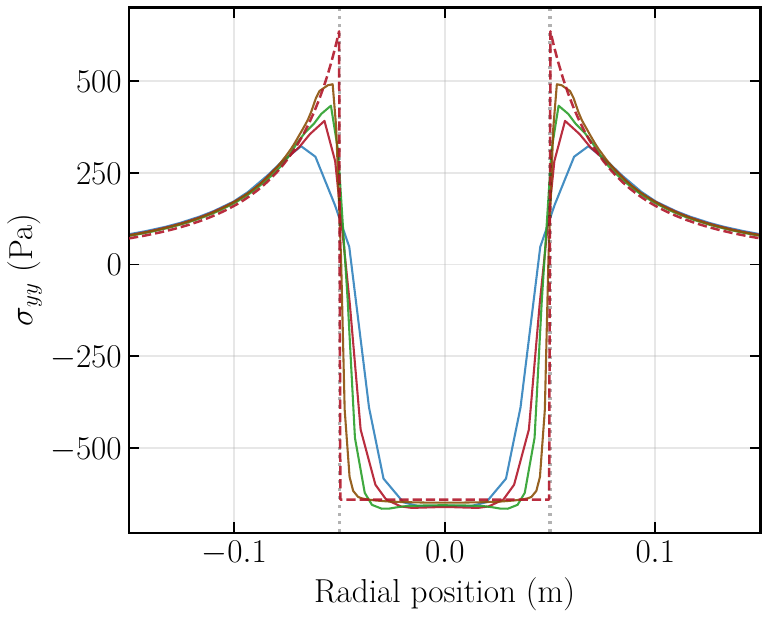}
        \caption{Single-domain, $E_\mathrm{in}/E_\mathrm{out}=2$}
        \label{fig:summary_yy_2_single}
    \end{subfigure}
    \par\vspace{0.12cm}

    \begin{subfigure}[b]{0.41\textwidth}
        \includegraphics[width=\textwidth]{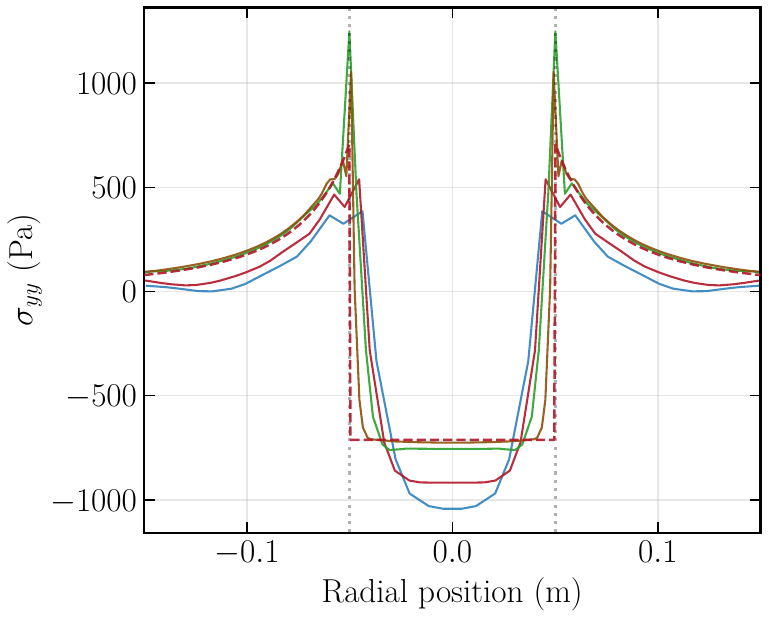}
        \caption{Dual-domain, $E_\mathrm{in}/E_\mathrm{out}=5$}
        \label{fig:summary_yy_5_double}
    \end{subfigure}\hfill
    \begin{subfigure}[b]{0.41\textwidth}
        \includegraphics[width=\textwidth]{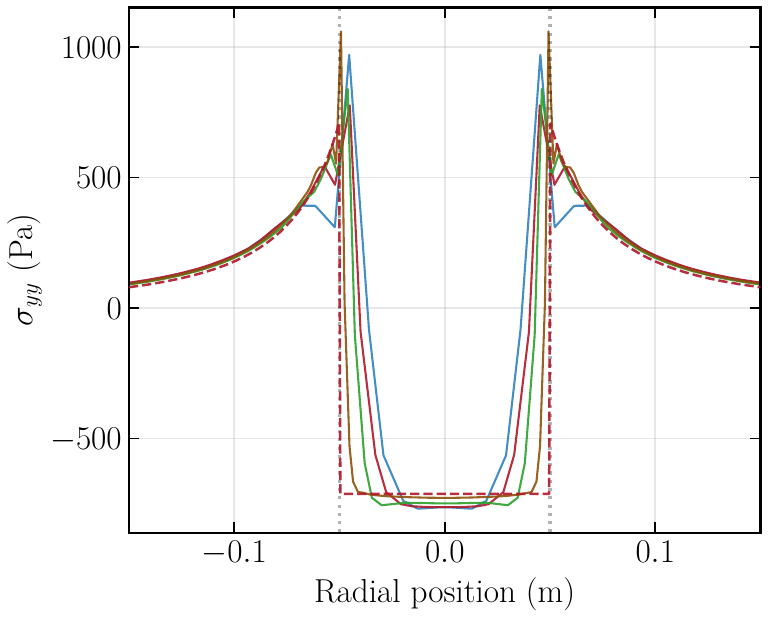}
        \caption{Single-domain, $E_\mathrm{in}/E_\mathrm{out}=5$}
        \label{fig:summary_yy_5_single}
    \end{subfigure}
    \caption{$\sigma_{yy}$ stress profile along a horizontal cross-section through the inclusion center at each refinement level, for three stiffness ratios ($E_\mathrm{in}/E_\mathrm{out} \in \{1, 2, 5\}$). Left column: our OS-MPM with dual domains; right column: single-domain MPM. Red dashed line: analytical solution.}
    \label{fig:summary_yy}
\end{figure}

The \emph{convergence} results under refinement  are plotted in Fig.~\ref{fig:summary_yy} the $\sigma_{yy}$, where stress profile is extracted along a horizontal cross-section through the inclusion center for all three stiffness ratios. The dual-domain and single-domain results at each refinement level are compared against the analytical solutions. In every case, the numerical profiles converge monotonically toward the analytical curve as the mesh is refined. OS-MPM tracks the single-domain convergence behavior closely, demonstrating that the Schwarz coupling does not degrade the asymptotic accuracy of the fine-domain solution even as the stiffness contrast increases to $E_{\mathrm{in}}/E_{\mathrm{out}} = 5$.


\begin{figure}[htbp]
    \centering
    \captionsetup[subfigure]{font=footnotesize}
    \includegraphics[width=0.4\textwidth]{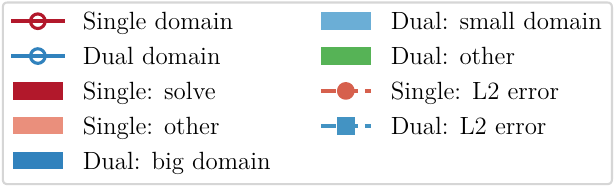}
    \par\vspace{0.08cm}
    \begin{subfigure}[b]{0.39\textwidth}
        \includegraphics[width=\textwidth]{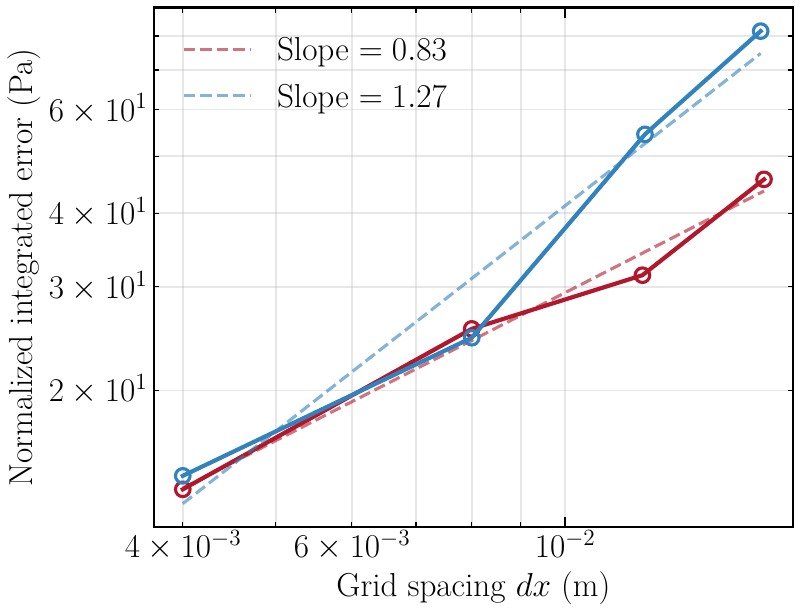}
        \caption{Error convergence, $E_\mathrm{in}/E_\mathrm{out}=1$}
        \label{fig:conv_1}
    \end{subfigure}\hfill
    \begin{subfigure}[b]{0.47\textwidth}
        \includegraphics[width=\textwidth]{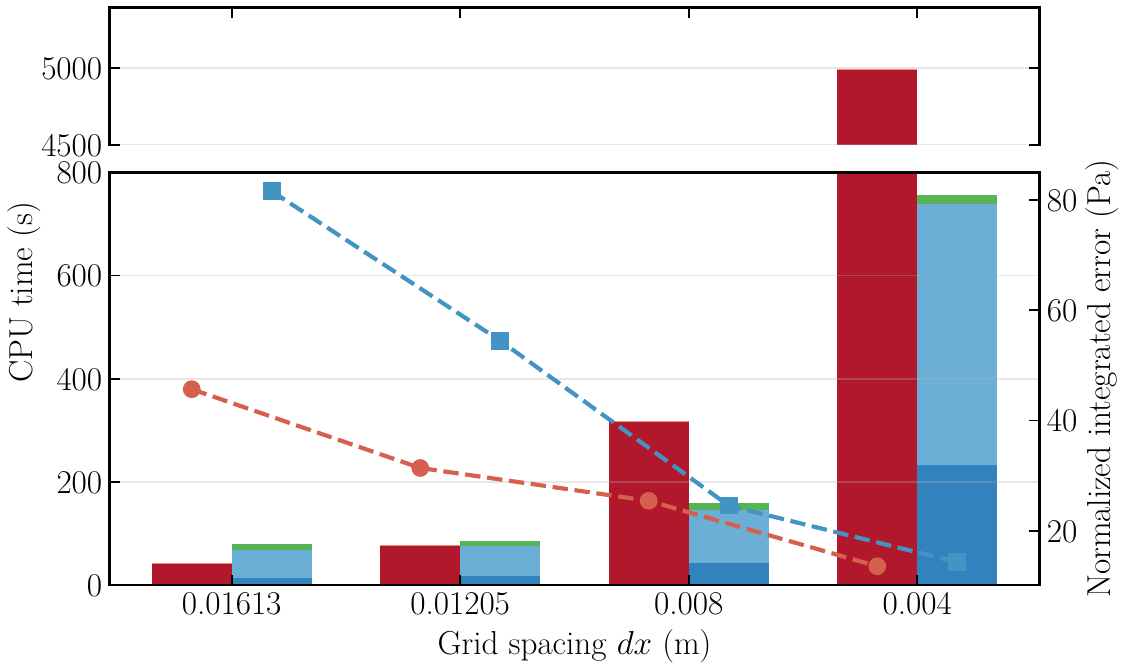}
        \caption{Computational efficiency, $E_\mathrm{in}/E_\mathrm{out}=1$}
        \label{fig:eff_1}
    \end{subfigure}
    \par\vspace{0.08cm}
    \begin{subfigure}[b]{0.39\textwidth}
        \includegraphics[width=\textwidth]{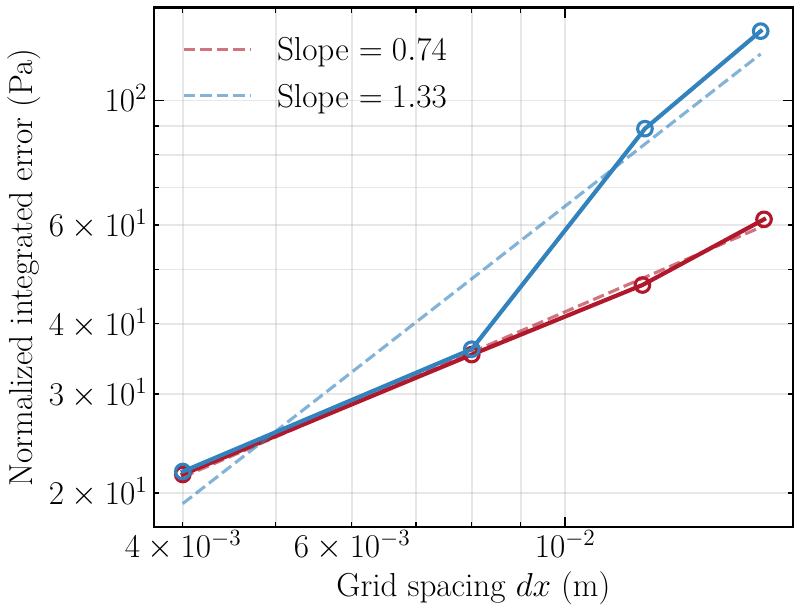}
        \caption{Error convergence, $E_\mathrm{in}/E_\mathrm{out}=2$}
        \label{fig:conv_2}
    \end{subfigure}\hfill
    \begin{subfigure}[b]{0.47\textwidth}
        \includegraphics[width=\textwidth]{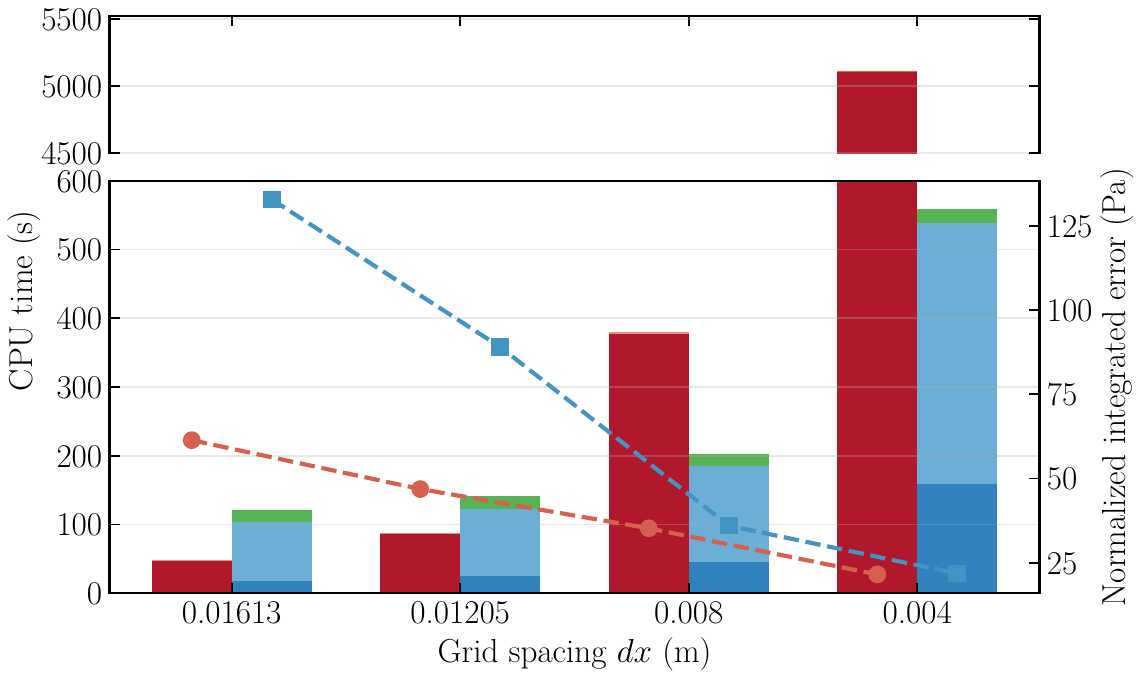}
        \caption{Computational efficiency, $E_\mathrm{in}/E_\mathrm{out}=2$}
        \label{fig:eff_2}
    \end{subfigure}
    \par\vspace{0.08cm}
    \begin{subfigure}[b]{0.39\textwidth}
        \includegraphics[width=\textwidth]{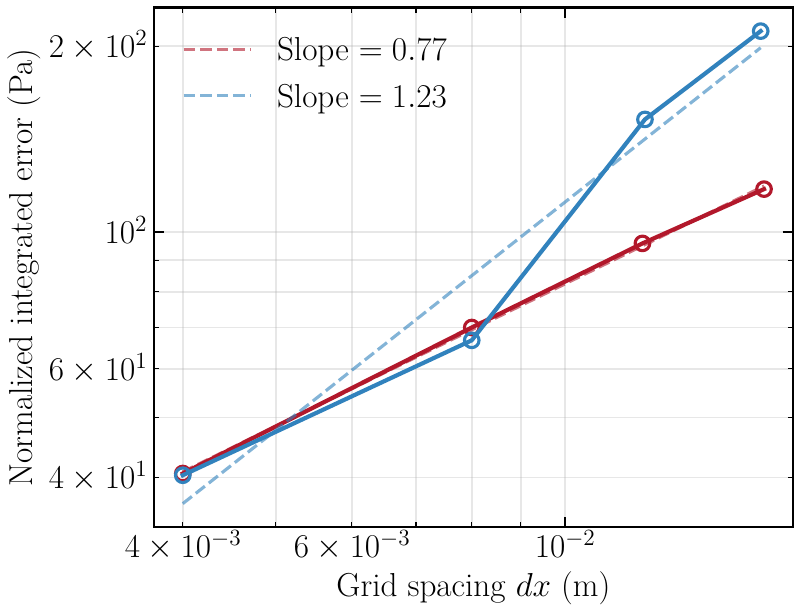}
        \caption{Error convergence, $E_\mathrm{in}/E_\mathrm{out}=5$}
        \label{fig:conv_5}
    \end{subfigure}\hfill
    \begin{subfigure}[b]{0.47\textwidth}
        \includegraphics[width=\textwidth]{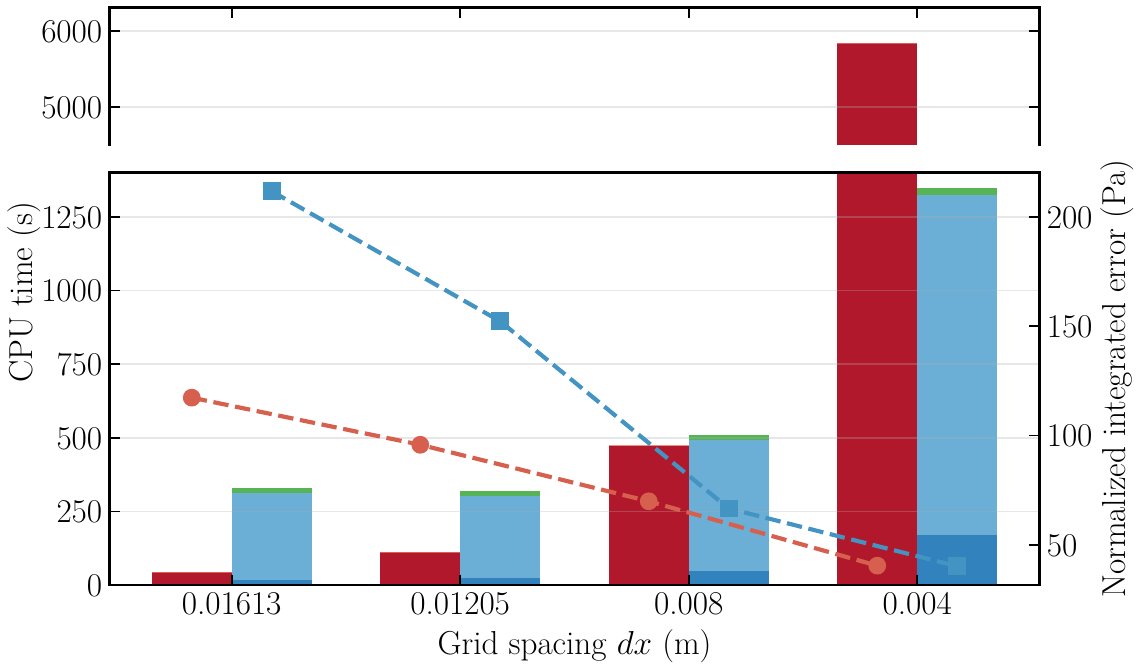}
        \caption{Computational efficiency, $E_\mathrm{in}/E_\mathrm{out}=5$}
        \label{fig:eff_5}
    \end{subfigure}
    \caption{Accuracy and efficiency comparison for the elastic inclusion problem ($E_\mathrm{in}/E_\mathrm{out} \in \{1,2,5\}$). Left column: $L_2$ stress error vs.\ fine-domain grid spacing. Right column: $L_2$ error vs.\ CPU time.}
    \label{fig:inclusion_efficiency}
\end{figure}

The \emph{error convergence and computational efficiency}, presented in
Fig.~\ref{fig:inclusion_efficiency}, quantify both the accuracy and the computational cost of the two solvers across all refinement levels and stiffness ratios. The left column of each row plots the $L_2$ error norm of the stress field against the fine-domain grid spacing; the right column plots the same error against CPU time. 

At the coarsest resolution the dual-domain error is marginally larger than the single-domain error. This is expected: the coarse domain $\Omega_B$ operates at half the spatial resolution, so its contribution to the global stress field is less accurate at the beginning of the refinement sequence. As the mesh is refined the gap closes rapidly, and at the finest tested resolution the dual-domain error is comparable to, or slightly below, the single-domain error in all three stiffness cases. Importantly, the convergence rate of OS-MPM is consistently faster than that of the single-domain method, indicating that targeted refinement of $\Omega_S$ is a more effective use of degrees of freedom.

The efficiency advantage is equally clear. At coarse resolutions the dual-domain solver carries a modest overhead from the alternating Schwarz iterations, so its CPU time is slightly higher than the single-domain reference. However, the computational cost of the single-domain solver grows much faster with refinement. In contrast, the cost of OS-MPM grows slowly since only the fine subdomain is refined. At the finest resolution tested, OS-MPM achieves speedups of $6.61\times$, $9.15\times$, and $4.34\times$ over the single-domain solver for stiffness ratios $E_{\mathrm{in}}/E_{\mathrm{out}} = 1$, $2$, and $5$, respectively. The increasing speedup with stiffness ratio reflects the fact that higher contrast concentrates the stress gradient more tightly near the inclusion boundary, making selective refinement of $\Omega_S$ even more beneficial.

\subsection{Three-Dimensional Showcase}
\label{subsec:exp_showcase_3d}
As a qualitative three-dimensional demonstration, we consider a foldable-display-inspired structure consisting of two stiff panels connected by a compliant hinge. This example is motivated by engineering systems in which deformation is intentionally localized within a narrow functional region, while the surrounding components are expected to remain comparatively rigid \cite{kim2023fabrication,zhao2022intrinsically}. Such configurations are well suited to the proposed OS-MPM framework, because high spatial and temporal resolution is primarily required near the hinge and transition zones, whereas the stiff panels can be represented at a coarser resolution without sacrificing the dominant deformation response.

The structure has initial dimensions $0.3 \text{cm}\times 9 \text{cm} \times 6\text{cm}$ in the $x$, $y$, and $z$ directions, respectively. The two straight panels are assigned a Young’s modulus $E=1.0\times10^2$ kPa, while the compliant hinge has $E=1.0$ kPa. Both regions use a Poisson’s ratio of $\nu=0.4$, and a linear transition zone of width 0.03 cm is introduced between the stiff and compliant regions. In the domain-decomposed model, the hinge and its neighboring transition region are resolved by the fine domain $\Omega_s$, while the stiff panels are covered by the coarse domain $\Omega_B$. The grid spacing is $h = 5.6\times10^{-3}\,\mathrm{cm}$, the time step is $\Delta t = 5\times 10^{-4}\,\mathrm{s}$, and each grid cell is initialized with $4$ particles.

\begin{figure}[htbp]
    \centering
    \begin{subfigure}[b]{0.48\textwidth}
        \centering
        \includegraphics[width=\textwidth]{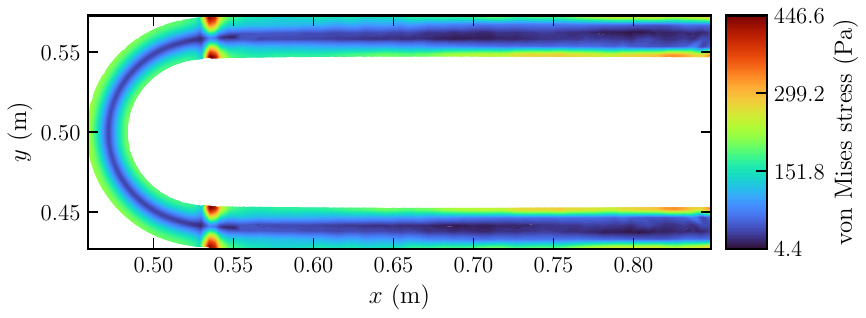}
    \end{subfigure}
    \hfill
    \begin{subfigure}[b]{0.48\textwidth}
        \centering
        \includegraphics[width=\textwidth]{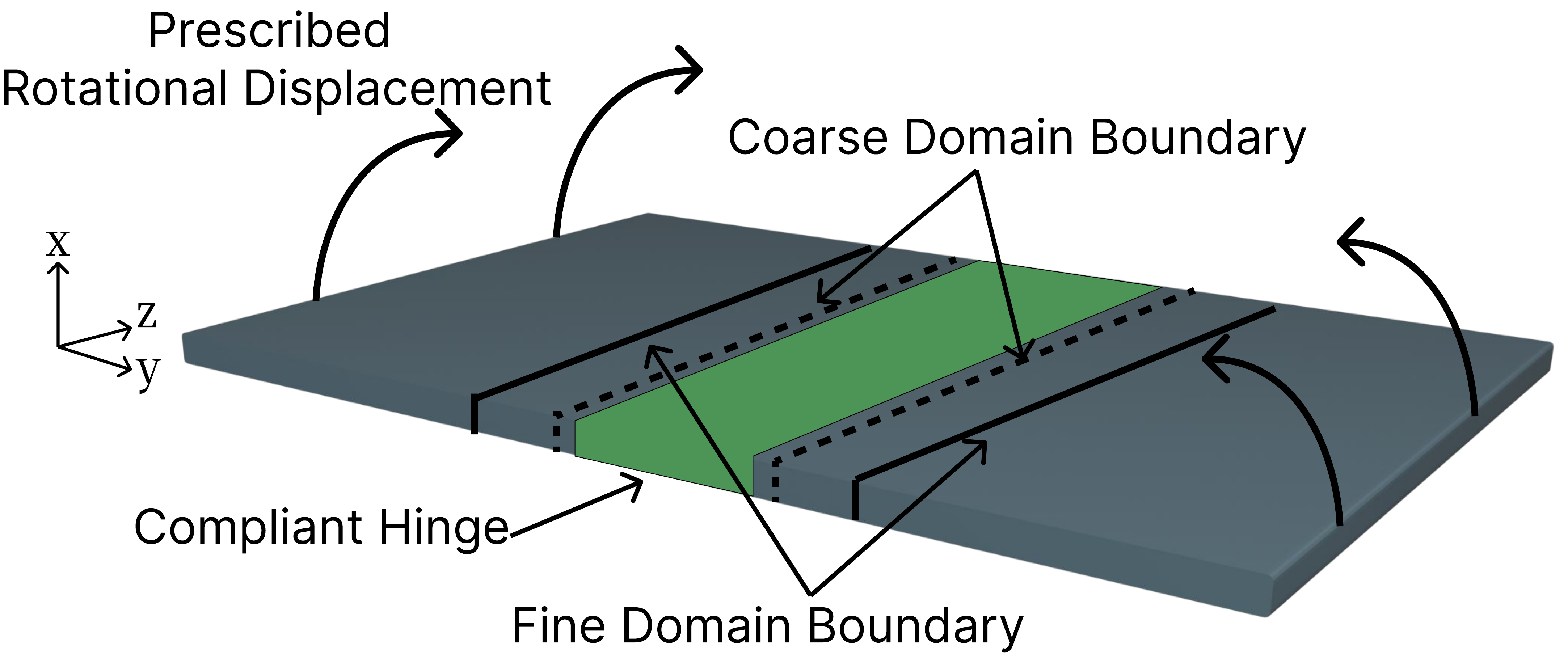}
    \end{subfigure}

    \vspace{0.08cm}

    \begin{subfigure}[b]{0.48\textwidth}
        \centering
        \includegraphics[width=\textwidth]{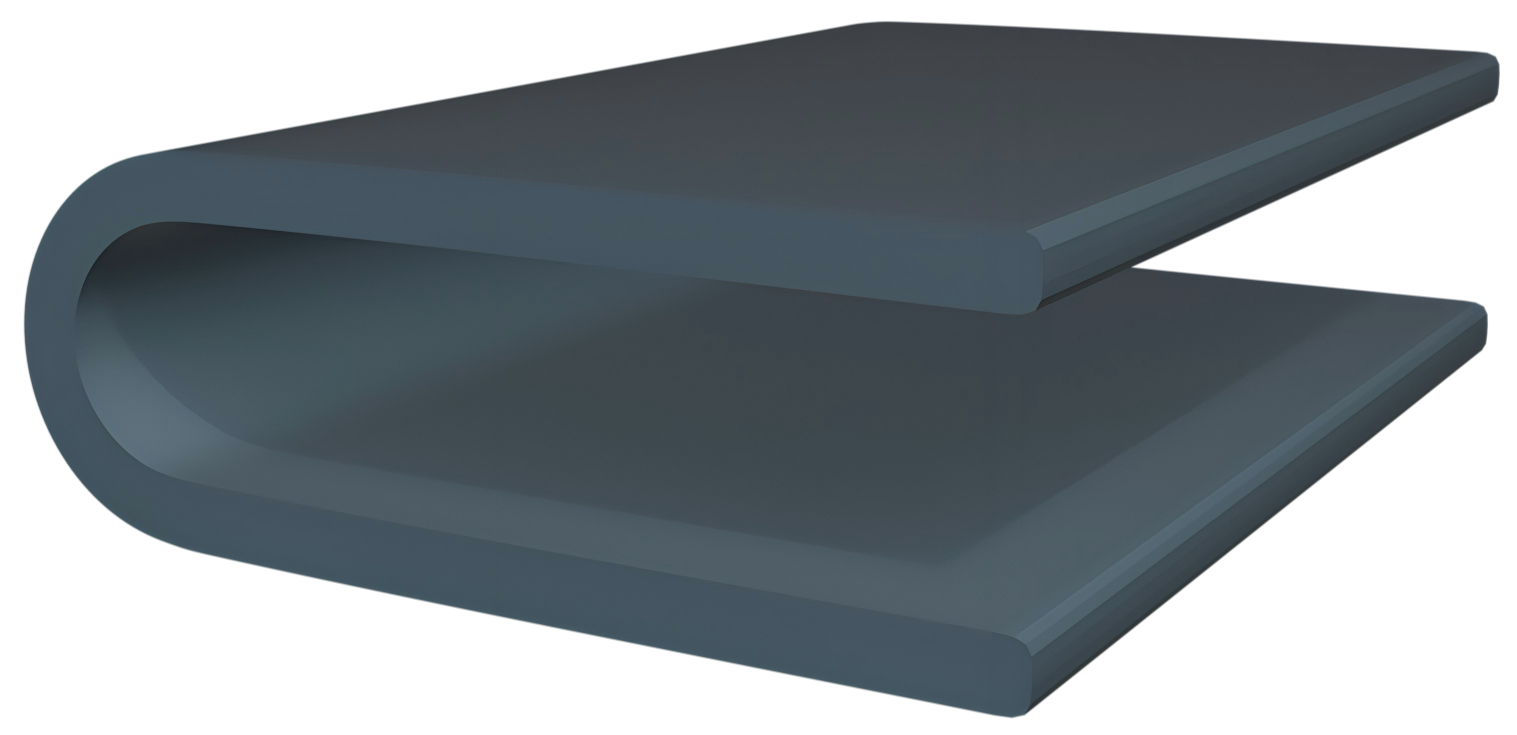}
    \end{subfigure}
    \hfill
    \begin{subfigure}[b]{0.48\textwidth}
        \centering
        \includegraphics[width=\textwidth]{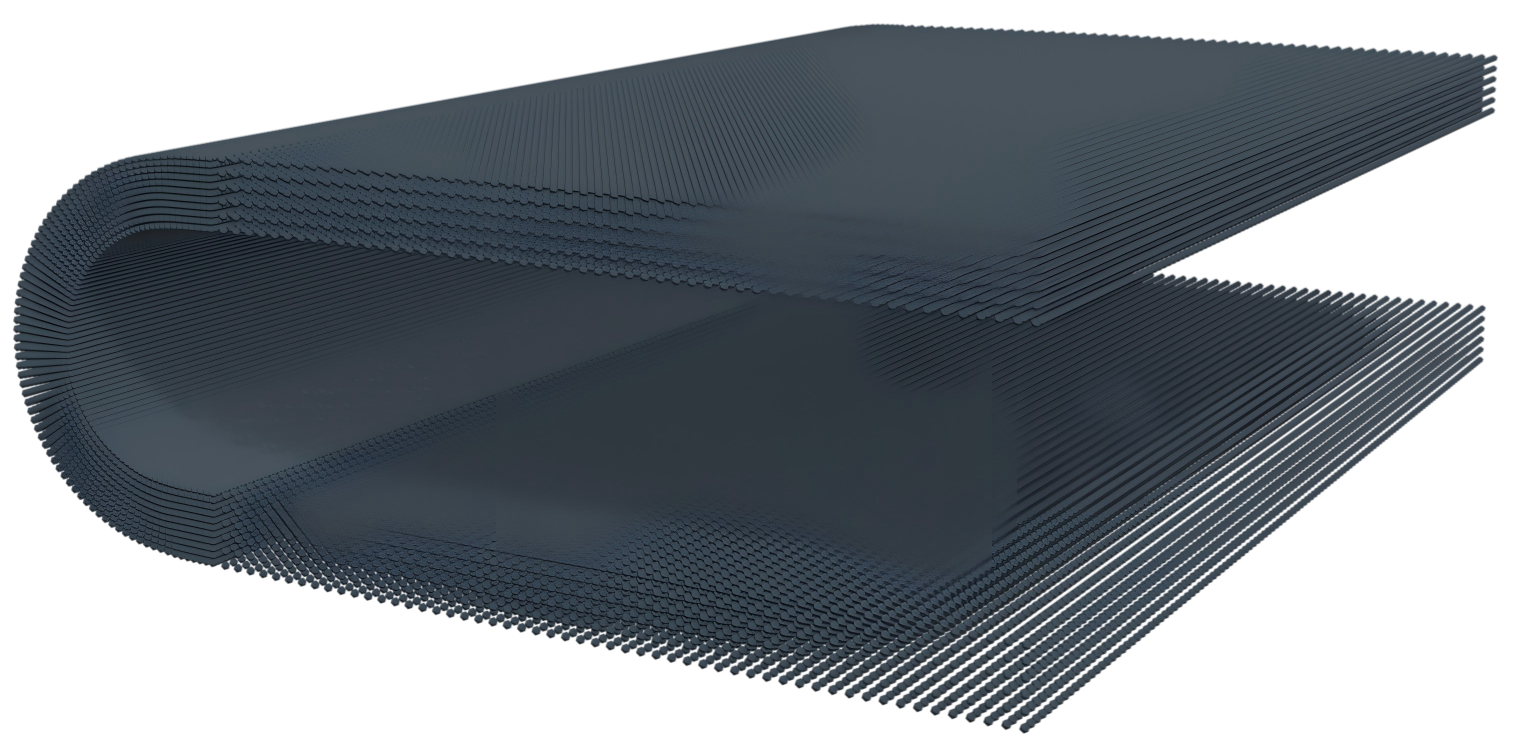}
    \end{subfigure}
    \caption{Three-dimensional foldable-display-inspired showcase. Top left: von Mises stress distribution at the cross-section $z=0.3$. Top right: rest configuration and experimental setup. Bottom: solid (left) and particle (right) visualization of the folded configuration.}
    \label{fig:showcase_3d}
\end{figure}

Figure~\ref{fig:showcase_3d} shows the stress distribution, rest configuration, folded shape, and particle distribution. The resulting deformation is smooth and symmetric, with bending localized primarily in the compliant hinge while the two panels retain their intended rigidity. The von Mises stress concentration along the hinge and near the material-transition interfaces is clearly captured, indicating that the coupled coarse–fine discretization can resolve the mechanically critical region without visible artifacts across the Schwarz interface. This feature is particularly relevant for foldable electronic devices, where stress concentration and repeated deformation around the hinge are closely related to durability and failure. Although the present example is limited to an elastic demonstration, the same framework provides a natural basis for future studies involving damage evolution, elastoplastic deformation, fatigue-like degradation, or interface failure in hinge-dominated structures.

%% file: sections/conclusion.tex
\section{Conclusion}
\label{sec:conclusion}





We have presented an overlapping Schwarz space-time refinement framework for the material point method (OS-MPM) that enables concurrent coarse-fine coupling over subdomains with heterogeneous spatial and temporal resolutions. The method retains standard MPM discretizations within each subdomain and realizes the coupling through an MPM-specific Schwarz iteration based on mass-weighted spatial projection, temporal interpolation, and fine-scale sub-cycling. In this way, local refinement is achieved without globally modified basis functions, special transition kernels, or saddle-point/penalty interface treatments, thereby preserving the modularity of the underlying MPM formulation.

The results show that the proposed framework preserves the accuracy and convergence characteristics of monolithic fine-resolution simulations while reducing computational cost when fine resolution is only needed locally. Across the benchmark problems considered, the method accurately captures strongly nonlinear deformation, localized contact, and sharp stress gradients associated with material heterogeneity. In particular, the elastic inclusion study demonstrates that OS-MPM can achieve comparable accuracy to single-domain fine simulations at substantially lower runtime, with speedups reaching 9.15 times in the reported cases. The successful three-dimensional showcase further highlights the generality of the approach.

These findings suggest that OS-MPM is a promising and practical strategy for local space-time refinement in MPM. It opens a path toward efficient simulation of multiscale solid mechanics problems in which localized features govern accuracy but global uniform refinement would be prohibitively expensive. Future developments will focus on adaptive domain placement, more general multi-subdomain configurations, and scalable parallel implementations for large-scale three-dimensional problems.

\section*{Acknowledgments}
We sincerely thank Qirui Fu for valuable discussion. Yupeng Jiang acknowledges the financial support from the International Research Training Group (IRTG) 2657, funded by the German Research Foundation (DFG) (Project-ID 433082294). Minchen Li acknowledges Carnegie Mellon University for the Junior Faculty Startup Fund and Genesis AI for the gift funding.

%% file: sections/appendix.tex



%

%



